\documentclass{article}

\begin{document}

\textheight=8.2in \textwidth=10in

\title{\LARGE \bf Interaction of bimodal fields with few-level atoms in cavities and traps }

\author{A. Messina, S. Maniscalco and A. Napoli \\ INFM, MURST and Dipartimento di Scienze Fisiche ed Astronomiche,\\ via Archirafi 36, 90123 Palermo, Italy \\ Tel/Fax: +39 91 6234243; E-mail: messina@fisica.unipa.it }
%\address{INFM, MURST and Dipartimento di Scienze Fisiche ed Astronomiche,\\ via Archirafi 36, %90123 Palermo, Italy \\ Tel/Fax: +39 91 6234243; E-mail: messina@fisica.unipa.it }

\maketitle

\begin{abstract}
The spectacular experimental results of the last few years in
cavity quantum electrodynamics and trapped ions research has led
to very high level laboratory performances. Such a stimulating
situation essentially stems from two decisive advancements. The
first is the invention of reliable protocols for the manipulation
of single atoms. The second is the ability to produce desired
bosonic environments on demand. These progresses have led to the
possibility of controlling the form of the coupling between
individual atoms and an arbitrary number of bosonic modes. As a
consequence, fundamental matter-radiation interaction models like,
for instance, the JC model and most of its numerous nonlinear
multiphoton generalizations, have been realized or simulated in
laboratory and their dynamical features have been tested more or
less in detail. This topical paper reviews the state of the art of
the theoretical investigations and of the experimental
observations concerning the dynamical features of the coupling
between single few-level atoms and two bosonic modes. In the
course of the paper we show that such a configuration provides an
excellent platform for investigating various quantum intermode
correlation effects tested or testable in the cavity quantum
electrodynamics  and trapped ion experimental realms. In
particular we discuss a  mode-mode correlation effect appearing in
the dynamics of a two-level atom quadratically coupled to  two
bosonic modes.  This effect, named parity effect, consists in a
high sensitivity to the evenness or oddness of the total number of
bosonic excitations.

\end{abstract}

\pagebreak

 \tableofcontents

\pagebreak

\section{Foreword}

The history of the basic ideas of quantum theory, is studded with
proposal of \textit{gedanken experiments}. By this term, broadly
speaking, one refers to rather idealized physical scenarios
accompanied by reasonings aimed at highlighting some specific
conceptual aspects abstracted from a difficult scientific problem,
with disregard of any technological limitation. One of the most
famous \textit{gedanken experiments} was proposed by
Schr\"{o}dinger to draw the attention  to the  paradoxical results
to which a blind application of quantum mechanics to the
macroworld may lead. This example is known as the Schr\"{o}dinger
cat paradox and today it represents the archetype of all the
physical situations wherein quantum superpositions of two
macroscopically distinguishable states are taken into
consideration. It is interesting to point out that some
\textit{gedanken experiments}, differently from the one involving
the unlucky cat, are not only conceptual constructions but are
also conceived as real, albeit impracticable, experimental
schemes.

The peculiar feature of a thought experiment of this kind is the
striking contrast between the simplicity of the ingredients
involved
 (a single atom, a single harmonic oscillator, a single
quantized electromagnetic mode, etc.) and the unavailability of
the necessary experimental setup.

Fortunately, owing to technological progress, experiments
considered unrealistic up to recently, can now be carried out. As
a result, various intriguing genuinely quantum effects traceable
back to the superposition principle, entanglement, quantum
interference, etc., are now within experimental reach. Experiments
of increasing difficultiy in cavity quantum electrodynamics (CQED)
 over the last years, have, for example, made it possible to
test fundamental radiation-matter interaction models involving
single atoms. Moreover, cavity field states possessing remarkable
nonclassical features have been generated and detected.

In the domain of trapped ions, on the other hand, a single
material quantum harmonic oscillator has been experimentally
realized and the possibility of tailoring at will its coupling
with electronic degrees of freedom has been practically
demonstrated. It is therefore to be expected that these  two
research areas will play a decisive role in resolving the basic
paradoxical aspects of quantum physics.

The scope of this paper is to review the state of the art of the
theoretical investigations and of the experimental observations
concerning the dynamical features of the coupling between single
few-level atoms and two bosonic modes. The physical situations
which we shall refer to, belong to the experimental domains of
cavity quantum electrodynamics and of trapped ion physics. In the
latter case we shall describe the physical properties of an ion
harmonically confined in a bidimensional Paul microtrap. The
motion of the center of mass of the ion is coupled to its internal
dynamics using appropriate laser beam configurations. The
electronic degrees of freedom of the ion are represented by
pseudo-spin operators whereas the motional ones by bosonic
creation and annihilation operators associated to the unperturbed
bidimensional harmonic motion of the ion.

The physical scenario relative to the problems we shall be faced
with in a CQED context, involves one or more two- or three-level
atoms interacting, one at a time, with two quantized
electromagnetic modes sustained by a high-$Q$ resonator. In this
case too, the dynamical properties of the system are investigated
using hamiltonian models characterized by the presence of bosonic
variables describing the two quantized electromagnetic modes and
of pseudo-spin atomic operators. Thus the two different research
areas share, at least, the mathematical language and possibly, in
addition, the potentiality for realizing, with the help of
different experimental setups, classes of simple but basic pseudo
spin-boson hamiltonian models.

It turns out that this is indeed true, even if, as we shall see,
the trapped ion context exhibits a greater practical feasibility
which suggests experimental investigations beyond the
possibilities of a typical CQED apparatus.

What are the motivations which justify a review of these physical
situations? In particular: what is the specific role of the two
bosonic modes?

A first answer is that the dynamics of several hamiltonian models
describing such systems is exactly treatable and, in most cases,
testable in the laboratory. A second more intriguing reason is
that investigating these systems is likely to shed light on basic
questions of quantum mechanics. The point to be appreciated is
indeed that, studying such systems, one has the opportunity to
induce entanglement and to control its evolution in a tripartite
physical system atom - mode 1 -mode 2 with  acceptable
experimental efficiency.

As we shall see, this entangled condition is essential for many
theoretical proposals and experimental studies, both in CQED and
trapped ion context, concerning the generation of a large class of
interesting entangled states of the two bosonic modes only. We
emphasize, in addition, that the presence of two modes in the
physical systems under scrutiny, can be successfully exploited
also to produce quantum target states of one mode only. In this
last application the other mode plays an auxiliary role, leading
to an effective improvement of the preparation protocol.

The quantum states of the bimodal subsystem is intrinsically
interesting because, for example, it allows the production of
bimodal states having nonclassical signatures. Of particular
relevance is also the fact that the preparation of some entangled
bimodal states yields the possibility of following the degradation
of  quantum coherence, under almost ideal conditions.

The importance of experimental observations of this kind might be
crucial to elucidate the nature of the decoherence mechanisms,
unavoidably plaguing the time evolution of any mesoscopic or
macroscopic physical system. Quite recently it has been possible
to follow in the laboratory the degradation  of mesoscopically
sized Schr\"{o}dinger cat like states advancing towards the
exciting goal of realizing a gedanken experiment.

We wish to emphasize that also the time evolution of the
tripartite system is of interest in its own, at least under the
aspect of bringing to light the mechanism governing the onset and
the time development of  quantum mode-mode correlation effects. In
this paper, for example, we shall review the so called parity
effect recently discovered in the dynamics of a two-level atom
quadratically coupled to  two bosonic modes. This intermode
correlation effect is related to the granularity of the energy of
a quantum harmonic oscillator. The occurrence of some interesting
consequences of the parity effect both in CQED and trapped ion
contexts, shall be presented and discussed, clarifying  in this
way, the motivation for reviewing together phenomena belonging to
these two different domains.

This paper is organized in two large sections. The first  is
devoted to systematically presenting the dynamical properties of
the coupling between few-level atoms and two quantized
electromagnetic modes. The second  explores, in the same spirit,
the physics  of an ion confined in a bidimensional Paul trap and
subjected to appropriate laser beam configurations.

\pagebreak

\section{Atoms in bimodal cavities }
 \subsection{Introduction}

Over the last thirty years many theoretical investigations have
been addressed toward the understanding of nonlinear effects
occurring in the dynamics of an atom in a high-Q cavity. The
interest toward this research area was mainly spurred by the large
amount of experiments revealing the appearance of subpoissonian
photon statistics \cite{Rempe,subpoissonian} among other
intringuing features of quantum radiation-matter interaction
\cite{Haroche1,varie,Brune96-1}. Both theoretical and experimental
activities have concentrated on trying to understand simple non
trivial models of quantum optics involving a single atom, regarded
as a few effective energy levels, and one or more near resonant
modes of the quantized electromagnetic field. The prototype of
such systems, proposed by Jaynes and Cummings in 1963 \cite{JC},
describes a two-level atom resonantly interacting with a
single-mode quantized field. When the Rotating Wave Approximation
(RWA) is made, the model becomes exactly solvable and its
dynamical features can be analytically brought to  light revealing remarkable properties. A review of the Jaynes and Cummings model (JCM) has been presented by Shore and Knight in 1993 \cite{revJC}.
The discovery of interesting aspects of the JCM as well as
the developments in CQED experiments involving single  Rydberg
atoms within
 single-mode cavities \cite{Meschede-Walther}, have stimulated an intense research devoted at
  highlighting and generalizing the original idea and physical scenario presented by Jaynes and Cummings.
In particular their work has been extended to include more elaborate interactions,
 such as multiphoton \cite{B-S81,Gerry88,Ashraf90} or intensity-dependent coupling \cite{Knight86,P-K90,B-J89,B-M-N94}, damping and
 dissipation \cite{Quang91,B-K86}, multiple atoms \cite{T-K69,H-R85,K-Q90}, multilevel atoms \cite{B-S84,K-S87} and multimode field \cite{P-S86,B-M}.
It is interesting to underline that historically many extensions
 of the JCM have been reported as examples of exactly treatable
 nonlinear matter-field interaction models and that some of them,
including the photon loss, provide realistic description of the
experiments.

After the experimental work of Brune et al \cite{Brune87}, which
have successfully realized a two-photon single mode maser operated
in a high-Q cavity, a large amount of experimental and theoretical
research has been done on the atomic two-photon transition
processes in a cavity. The  experiment successfully performed at
the ENS laboratory, represents an important milestone
 in quantum optics because it has practically realized a two-photon cascade micromaser. Under the stimulus of the success of the experiment of Brune et al., other schemes for a two-mode
two-photon maser have been presented \cite{Maia-Nieto}.

More recently, much attention has been devoted to the generation
of entangled states of two modes of two spatially separated
cavities \cite{two-cavities}. Apart from their intrinsic
theoretical interest,  these states play an important role in the
framework of the new field of quantum communication. For example,
a method for generating a maximally entangled state  of two
cavities has been reported  as an intermediate step in the
teleportation procedure proposed by Davidovich et al
\cite{Davidovich2}. Recently, Zubairy et al \cite{Zubairy} have
shown the possibility of teleporting a radiation field state from
a cavity to another one, provided that the generation of a
two-resonator entangled state for fixed number of photons inside
the two cavities, is feasible. Notwithstanding the relevance of
this topic, the study of protocols aimed at entangling two
separate single mode cavities, goes beyond the scope of this
topical review.

Section II of this paper is devoted to
the interaction between bimodal cavities and few level atoms. In particular
in Section II B a brief review of the JCM is presented whereas in Section II C its generalization to a bimodal interaction is discussed putting
into evidence the interest stemming from the consideration of two modes instead of one.
In Section II D a short discussion on two-photon micromaser is presented underlining its difference with respect to the one-photon micromaser.
In Section II E interesting nonclassical dynamical
effects of two-mode JC models are reported both in correspondence to the nondegenerate
case $(\omega_1 \not= \omega_2)$ and in the degenerate
one $(\omega_1 = \omega_2)$ when one-photon processes are assumed.
In Section II F we concentrate our attention on nonlinear
interaction models for which the coupling mechanism between the
atom and the bimodal cavity is based on two photon processes. In
particular Section II F 1 is devoted to the so called Raman
coupled model characterized by the fact that the net gain or loss
of photons is zero in any process. The models adopted in Sections
II F 2 -II F 4  are such that the respective basic processes
involve the net gain or loss of a pair of photons having or not
the same energy. Finally, in the last section (section II G) of
this first part, several schemes aimed at generating nonclassical
states in a bimodal high-Q microcavity are reported.

\subsection{A brief review of the JCM}

In the spirit of this topical review, we start devoting this section to a
brief discussion of the JCM being it the simplest model describing the interaction between a single few-level atom and the quantized cavity field.
This model has been extensively studied by many authors and numerous non-classical effects have been predicted and in some cases verified in laboratory \cite{Rempe,subpoissonian,Haroche1,varie,Brune96-1,JC,Meschede-Walther}.
The JCM consists in a single two-level atom, whose two states are denoted by $\vert + \rangle$ and $\vert - \rangle$,
coupled to a quantized single mode of a cavity field having a frequency $\omega$ close to the transition frequency $\omega_0$ of the atom.
The effective hamiltonian describing such a physical system, in the Rotating Wave Approximation, can be written in the form $(\hbar=1)$:
\begin{equation}
\hat{H}_{JC}=\omega_0\hat{S}_z+\omega \hat{a}^{\dag}\hat{a}
+[\lambda \hat{a}^{\dag}\hat{S}_-+h.c.]
\end{equation}
where $\hat{a}$ and $\hat{a}^{\dag}$ denote the
annihilation and creation operator  of the cavity mode whereas
the atomic degrees of fredom are described by the pseudospin operators $\hat{S}_z$ and $\hat{S}_{\pm}$  satisfying the following relations
\begin{equation}
2\hat{S}_z
\vert \pm \rangle= \pm \vert \pm \rangle\;, \;\;\;\;\; \hat{S}_{\pm} \vert
\pm \rangle=0 \;,\;\;\;\; \hat{S}_{\pm}\vert \mp \rangle=\vert \pm
\rangle
\label{pseudospin}
\end{equation}
The coupling constant $\lambda$ depends on cavity volume, dipole strenghts and mode frequency \cite{revJC,Cohen}.
In the JCM the familiar Rabi oscillations characterizing the time evolution of the atomic inversion $S_z$, are affected by the initial distribution of photon numbers \cite{colrev}. It has indeed been demonstrated that the interference between all the possible Rabi frequencies involved in the dynamics of the system, causes a dephasing or \textit{collapse} of the Rabi oscillations. Moreover, the discrete nature of the photon number distribution leads to a partially rephasing or \textit{revival} of such oscillations. The existence of this revival property has been experimentally observed and, being related to the discretness
of the field excitations, provides direct evidence for the truly
quantum nature of radiation \cite{Brune96-1}.

\subsection{Two-mode Jaynes Cummings models}

A natural generalization of the JCM consists in considering a single atom interacting with two modes of the quantized cavity field.
When such a physical situation is considered, several effective models can be constructed depending on the interaction mechanism characterizing the specific coupling.
In particular, the Hamiltonian model describing the atom - two cavity modes coupling, assumes different forms depending on whether one-photon or two-photon processes are considered.
In this last case the energy exchange between the atom and the bimodal cavity is mediated by virtual levels \cite{Shore} not explicitely involved in the effective hamiltonian.
In order to better clarify this point, consider one of the method followed by many authors for deriving effective
models, namely the so-called \lq\lq adiabatic elimination of virtual
levels procedure\rq\rq  \cite{Alsing,Puri-Bullough,G-E1,J-P1,Cardimona}. Its starting point consists in writing the full
microscopic Hamiltonian in the dipole approximation and neglecting
all the counter-rotating contributions in accordance with the RWA.
Then the Heisenberg equations  of the field and  atomic operators
in the interaction picture are considered. If the population in a
given atomic level is relatively much smaller than the other ones,
then, formally integrating the relative Heisenberg equations, this
level, called virtual atomic level \cite{Shore},  can be adiabatically
eliminated.

It is important to underline that the elimination of these virtual or intermediate states in the construction of two-photon effective models, leads to Stark shifts of atomic energy levels. As we will see, these shifts are proportional to photon numbers and  act as intensity-dependent detuning.

The interest toward two-mode JCM
originates from the several remarkable features
characterizing their dynamics. In our opinion one of the most
interesting property of the two-mode JC models is that they
provide the possibility to use one mode to modulate, amplify or
control the other one. As we will show later, this property turns
out to be very useful in the production of nonclassical  states of
a  single cavity mode. Another relevant aspect characterizing such
kind of systems lies in their wide potential tunability  since
oscillation or amplification can occur at any two frequencies
$\omega_1$ and $\omega_2$, with eventually $\omega_1=\omega_2$,
such that $\omega_1+\omega_2$ matches the atomic transition
frequency.
Before concluding this section we wish to point out that in order to describe two-mode JC models, we will denote by $\hat{a}_i$ and $\hat{a}_i^{\dag}$ $(i=1,2)$ the annihilation and creation operators respectively of the $i-$th cavity mode. Moreover, when the atom is effectively describable as a two-level atom the
pseudospin operators $\hat{S}_z$ and $\hat{S}_{\pm}$, such that the condition (\ref{pseudospin}) are satisfied,
will be used.

\subsection{Two-photon micromaser}

Micromasers are quantum oscillators operating on a Rydberg atomic transitions, namely corresponding to high principal quantum numbers.
In a micromaser an atomic beam is injected in a high-$Q$ microwave cavity. The atomic injection rate is low enough to ensure that there is just a single atom inside the cavity at one time.
The experimental realization of these devices provides a very simple way to test fundamental predictions related to radiation-matter interaction.
In particular, when the basic interaction process between each atom and the cavity field involves the emission or absorbtion of one photon per atomic transition (one-atom micromaser), the micromaser provides a near-ideal realization of the problem firstly theoretically treated by Jaynes and Cummings.
These devices also allow the operation of two-photon oscillators involving a degenerate or nondegenerate two photon transitions between two atomic levels of the same parity.
These two-photon micromasers present additional interesting features with respect to the one-photon micromaser, which are connected to the starting up of the system.
Classically, the zero-field state is always stable so that an injected field is required in order to build up the oscillation.
However, quantum fluctuations may drive the occurrence of a vacuum state instability, thus inducing the system to begin the oscillation.
The two-photon micromaser indeed spontaneously starts without any triggering required.
Moreover the two-photon micromaser provides a clean setup for studying the two photon processes.
A two-photon micromaser was realized by Brune et al. \cite{tpm} using a niobium superconducting cavity with $Q\sim 10^8$, resonant at $\omega\sim68 GHz$. They in particular exploited a two-photon degenerate transition in Rubidium atoms.

When the two photons are emitted into different modes of the cavity (nondegenerate two-photon transition) important features of the two-photon process, such as for example the correlation between the two photons emitted in two different modes, can be studied.

It is important to emphasize that the mere fact
that the atomic system emits or absorbs photon pairs leads to a field statistics very different from the one of the ordinary
laser or maser fields and makes it a privileged candidate
for the generation of squeezed light.
In such systems the Rydberg atom plays a role similar to a coupler
which correlates the two modes of light changing the photon
statistics inside the cavity.

The main novelty of the
nondegenerate two-photon micromaser (two different cavity modes of
frequencies $\omega_1$ and $\omega_2$ respectively with $\omega_1
\not= \omega_2$), with respect to the degenerate one ($\omega_1 =
\omega_2$), is the strong correlation between the two different
modes, which may produce a 50 \% squeezing  in the difference of
intensities at steady state \cite{Maia-Nieto}.

\subsection{One-photon processes }

In this section we consider the coupling between few atoms and two
modes of a high-$Q$ cavity field assuming that the interaction
mechanism  relies on single-photon processes.
In particular in the subsection II E 1 we focus on the case of an
effective three-level atom and two nondegenerate modes of a single
cavity whereas in subsection II E 2 we
assume that the two cavity
modes are degenerate and that the atom can be effectively
represented as a two-level system.

\subsubsection{Three-level atom interacting with a bimodal cavity}

A natural extension of the JCM  consists of a three-level atom
coupled to one or two modes of a high-$Q$ cavity field. This
generalization allows, for example, the investigation of physical
phenomena associated with two-photon processes. A general
treatment of the theory concerning such  systems has been
presented in  \cite{Xiao-shen Li,Zhen-don Liu} where arbitrary
detunings are considered.
Denoting the three atomic levels by $\vert 0 \rangle$, $\vert 1
\rangle$ and $\vert 2 \rangle$ and limiting our discussion to the
electronic dipole transition $\vert 0 \rangle$-$\vert 1 \rangle$
and $\vert 1 \rangle$-$\vert 2 \rangle$, there are three possible distinct
atomic level configurations as shown in fig.~\ref{fig1}.
 Following  \cite{Yoo-Eberly},  these
configurations are called $\Lambda$, $\Xi$ and $V$. In this
context the transition $\vert 0 \rangle$-$\vert 2 \rangle$ is
forbidden.

When the  interaction between the atomic system and the two cavity
modes is considered, it is interesting to study the effects of
intermode field correlations on the dynamics of the atomic system
and viceversa. This particular aspect of the problem forms the
basis of the paper of Lai et al \cite{Lai}. In order to
investigate on the occurrence of dynamical effects related to the
presence of initial  correlation in the physical system, the
authors consider an SU(2) coherent state \cite{SU2}
as an example of a two-mode
correlated state and a three-level atom in the $\Lambda$
configuration as atomic transition sequence. The SU(2) coherent
state has the property that, indicating by $m$ the maximum
possible number of excitations in each mode, if $n$ photons are
observed in one mode then $(m-n)$ photons will be observed in the
other one. This means that the SU(2) coherent state is an
eigenstate of the total excitation number operator
$\hat{a}_1^{\dag}\hat{a}_1+\hat{a}_2^{\dag}\hat{a}_2$ relative to
the eigenvalue $m$.
The three-level atomic system consists of two dipole allowed
transitions, $\vert 0 \rangle$-$\vert 1 \rangle$ and $\vert 1
\rangle$-$\vert 2 \rangle$ respectively (see fig.~\ref{fig1}(a)),
resonant with two
different modes of the quantized radiation field. The interaction
Hamiltonian describing the coupled system, in the RWA, can be
written down as

\begin{equation}\label{hlambda}
 \hat{H}_I^{\Lambda}=g_1(\hat{a}_1^{\dag}\vert 0 \rangle \langle 1 \vert
  +h.c.)+g_2(\hat{a}_2^{\dag}\vert 2 \rangle \langle 1 \vert
  +h.c.)
\end{equation}
where $g_1$ and $g_2$ are the coupling constants relative to the two different channels $\vert 0 \rangle-\vert 1\rangle$ and $\vert 2\rangle-\vert 1\rangle$ respectively.
 The particular form of initial field correlation imposed by the authors to the
two cavity modes, leads to generalized two-photon Rabi frequencies
that are independent of the total photon number $m$, provided that
the atom-field coupling constants are set equal. Thus, under this
condition, the atomic population probability exhibits purely
periodic behaviour. Moreover the revival times, that is the time instants
at which the oscillations revive, are much shorter than the
ones exhibited assuming the field in the two-mode uncorrelated
binomial state. This behaviour is of relevance in view of the fact
that the photon statistics of the individual modes are the same in
the two cases considered by the authors.

The $\Lambda$ configuration is also exploited in ref. \cite{Zheng}.
In this paper the authors define the slowly varying complex amplitudes of the two-mode field as following:
\begin{equation}
\hat{d}_1=\frac{1}{2^{\frac{3}{2}}}\sum_{j=1,2}{(\hat{a}_je^{i\omega_jt}+
\hat{a}^{\dag}_je^{-i\omega_jt})} \;\;\;\;\;
\hat{d}_2=\frac{i}{2^{\frac{3}{2}}}\sum_{j=1,2}{(\hat{a}_je^{i\omega_jt}-
\hat{a}^{\dag}_je^{-i\omega_jt})}
\end{equation}
where $\omega_j$ is the frequency of the $j-$th mode. By definition $[\hat{d}_1,\hat{d}_2]=\frac{i}{2}$ so that squeezing of the two-mode field occurs whenever one of the amplitude operators satisfies the condition
\begin{equation}
<(\Delta \hat{d}_j)^2> < \frac{1}{4}
\end{equation}
In particular  the authors demonstrate the occurrence of two-mode squeezing  starting from the two cavity modes prepared in a coherent state.
Other interesting phenomena
are found as, for example, the periodic recovery of squeezed states
caused by the involved two-photon processes.

The interaction between a three-level atom and two different
cavity modes has been shown to produce also normal as well as
higher-order squeezing. In ref. \cite{Mahran}, M.H. Mahran studies the interaction of a three-level
V-shaped atom (see fig.~\ref{fig1}(c)) with two quantized electromagnetic modes discussing the possibility of occurrence of squeezing.
Following the definition proposed by Loudon and Knight
\cite{Knight-Loudon}, the author has calculated the two-mode
squeezing when the atom is initially prepared in its ground state.
The results obtained by Mahran brought to  light the dependence of
the two-mode squeezing on the initial mean photon number. In
particular the observed amount of squeezing decreases as the mean
value of photons initially present in the two modes increases.
Finally the author considers the sum
squeezing, as defined by Hillery \cite{Hillery}, for an atom
initially prepared in its ground state, proving that such a
quantity is smaller than the amplitude-squared squeezing.

The ladder or $\Xi$ configuration, displayed in fig.~\ref{fig1}(b)
is investigated in  \cite{Poizat}  bringing to light that it can
produce very good quantum nondemolition measurement performances.

Phase properties of the field in the two-mode three-level problem
are investigated in  \cite{Ho}. In particular, using the Hermitian
phase formalism of Pegg and Barnett \cite{Pegg}, phase properties
of each individual mode as well as their joint probability
distribution and correlation function are studied. The authors
show that phase properties of the field reflect the collapses and
revivals of the atomic level occupation probabilities except for
the Raman scattering in correspondence of which the collapses and
revival of the atomic population can be completely decorrelated
from the phase of the field.

Very recently \cite{Arun} the
master equation for the reduced density operator of the field in
the two-mode micromaser pumped by ultracold $\Lambda$-type
three-level atoms has been derived and the steady state photon
probability distribution under the principle of detailed balance
has been evaluated. Analogously to the definition given by Scully
et al. \cite{Scully}, the
system now described is called two-mode \textit{mazer }(microwave
amplification by the $z$ motion induced emission of radiation).
When indeed the micromaser is pumped by laser cooled atoms,
quantization of atomic external motion becomes necessary.
The interesting result obtained by the authors is that the degree
of anticorrelation between the cavity modes increases when the
micromaser is pumped by ultracold atoms instead of fast atoms.
This leads to much stronger sub-Poissonian statistics for each
mode.

\subsubsection{Degenerate one-photon two-mode  model}

In this section we focus our attention on the
one-photon interaction
between a single two-level atom and two degenerate modes of a lossless
cavity described with the help of the following Hamiltonian model \cite{BM1}:
\begin{equation}\label{hgiovanni}
\hat{H}_d=\omega \sum_{\mu=1,2}{\hat{a}_{\mu}^{\dag}\hat{a}_{\mu}}+ \omega_0 \hat{S}_z+
\sum_{\mu=1,2}{g_{\mu}(e^{-i \phi_{\mu}}\hat{a}_{\mu}^{\dag}\hat{S}_-+h.c.)}
\end{equation}
This model conserves the total excitation number
$\hat{N}=\hat{a}_1^{\dag}\hat{a}_1+\hat{a}_2^{\dag}\hat{a}_2+\hat{S}_z+\frac{1}{2}$. Moreover it possesses the purely bosonic constant of
motion $C=\frac{g_2^2}{g_1^2+g_2^2}\hat{a}_1^{\dag}\hat{a}_1+
\frac{g_1^2}{g_1^2+g_2^2}\hat{a}_2^{\dag}\hat{a}_2
-\frac{g_1g_2}{g_1^2+g_2^2}(\hat{a}_1^{\dag}\hat{a}_2
e^{i(\phi_2-\phi_1)}+h.c.)$.
The existence of these two constants of motion leads to the
possibility of unitarily transforming $\hat{H}_d$ into a system
consisting of an one mode JC subsystem and a decoupled other new
bosonic mode. Thus the dynamics of the model given by equation\
(\ref{hgiovanni}), may be exactly solved starting from an
arbitrary initial condition.

When one mode, say the first, is prepared in a Fock state with
$n\gg 1$ excitations and the other one is empty, the field
dynamics is dominated by a set of $n$ incommensurate Rabi
frequencies leading to the occurrence of oscillatory net exchanges
of a large number of photons between the two modes. These
oscillations of the field populations display amplitude decay and,
over a longer time scale, exhibit revivals and collapses. This
behaviour is displayed in fig.~\ref{fig2} where we have in particular
chosen $n=20$.
Analogous phenomena take place in the dynamics of the system when
the initially populated mode is prepared in a coherent state\cite{Benivegna}.  The reaction of the atomic dipole on the field
and the granularity of the field itself, are at the physical
origin of the highly nonclassical regime attained by the system
under scrutiny. In  \cite{Benivegna} the role played by the
atom-cavity field entanglement is brought to light and related to
the occurrence of a cooperative intermode effect which is shown to
be incompatible with a neoclassical approach.

\subsection{Nonlinear interaction models}

Over the last two decades the interest toward
the dynamics of nonlinear models effectively describing the interaction
between an atom and two quantized modes of a high-$Q$ cavity, has grown up in a significative way. With
the emergence of the two-mode two-photon micromaser of Laughlin
and Swain \cite{Laughlin-Swain}, it has become possible to experimentally observe
certain nonlinear effects including field squeezing and emission
spectra.
Both theoretical and experimental activities concentrate on trying
to understand simple non trivial models of quantum optics
involving single effective two-level atoms and two independent
modes of the radiation field, having or not the same frequency,
coupled by means of two-photon processes.
In this section we focus our attention on nonlinear two-mode
models, bringing to  light the main features characterizing their
relative dynamics.

\subsubsection{Raman coupled model }

Let's  consider two quantized cavity modes of frequency $\omega_1$
and $\omega_2 (< \omega_1)$ respectively and  a three-level
Rydberg atom whose energy diagram is given by fig.~\ref{fig3}.
Following Gerry and Eberly \cite{G-E1}, the level 2 can
be adiabatically eliminated provided that the condition of large
detuning $\Delta \gg E_3-E_1$ is satisfied.

Under this hypothesis the effective hamiltonian describing the
system under scrutiny can be written down in the following form:
\begin{equation}\label{hraman}
 \hat{H}_R=\hat{H}_0^R+\hat{H}_I^R
\end{equation}
with
\begin{equation}\label{hraman01}
  \hat{H}_0^R=\omega_0\hat{S}_z+\omega_1\hat{a}_1^{\dag}\hat{a}_1 +\omega_2\hat{a}_2^{\dag}\hat{a}_2; \;\;\;\;\;\;\;\;
  \hat{H}_I^R=g_R\hat{a}_1\hat{a}_2^{\dag}\hat{S}_++h.c.
\end{equation}
The resulting model therefore consists of
two-nondegenerate \lq \lq ground\rq \rq states connected by
zero-photon processes where a photon absorbed in one mode is
emitted in the other one with the atom making transitions through
the virtual state $\vert 2 \rangle$\cite{G-E1}.

The system under scrutiny, effectively described by $\hat{H}_R$, may be
interpreted as a cavity version of Raman scattering in which mode
1 is the pump field, mode 2 is the Stokes field and the
anti-Stokes mode is eliminated by setting a cavity off-resonance
condition \cite{Deb}. As a direct consequence of the fact that we are dealing with zero photon process, the total photon number operator defined as $\hat{N}=\hat{a}_1^{\dag }\hat{a}_1+\hat{a}_2^{\dag }\hat{a}_2$ is a constant of motion \cite{G-E1}.

The two-mode Raman coupled model considered in this section, has
been extensively investigated showing the existence of collapses
and revival phenomena \cite{G-E1}, resonance fluorescence
\cite{Ray} , population trapping  \cite{Deb} and several
nonclassical properties of the electromagnetic field
 such as the appearing
of antibunching light, violation of the Cauchy-Schwartz inequality
and squeezing \cite{G-E1,G-H1}.
 As far as
the  collapse and revival phenomena of the Rabi oscillations, it
has been proved that, differently from the JC model, in this case
it can be periodic \cite{G-E1,G-H1}. In particular, assuming coherent
states for both modes, the atomic inversion between the two
nondegenerate ground states exhibits periodic revivals but with
many secondary revivals as well. These last revivals become less
significant for higher initial average photon numbers. The
physical origin of such collapse and revival patterns was
partially brought to  light in \cite{G-H1} but it has been more
studied, at least for the case of equal average photon numbers in
both modes, by Cardimona et al. \cite{Cardimona}. As far as
nonclassical properties of the electromagnetic field are
concerned, Mahran \cite{Mahran2}, has, for example, studied the
production of squeezing.

The Hamiltonian model \ (\ref{hraman})-\ (\ref{hraman01}) adopted
in  \cite{G-E1,Cardimona,Deb,Mahran2} neglects the Stark shift
terms arising from the adiabatic elimination of a virtual level.
Relaxing this approximation  implies indeed that the resulting
two-level system is no longer two-photon resonant but it is
detuned by an intensity-dependent amount. In other words the
unperturbed atomic hamiltonian $\omega_0\hat{S}_z$ appearing in
$H_0^R$, has to be substituted by
\begin{equation}
(E_3+\frac{g_2^2}{\Delta} \hat{a}_2^{\dag}\hat{a}_2)\vert 3 \rangle \langle 3 \vert +
(E_1+\frac{g_1^2}{\Delta} \hat{a}_1^{\dag}\hat{a}_1)\vert 1 \rangle \langle 1 \vert
\end{equation}
where $g_i$ $(i=1,2)$ is the atom $i-th$ mode coupling constant.
Buzek
and Knight \cite{B-K} have shown that this substitution may lead to
quite different evolutions even for low-intensity fields.
 The effects of the Stark shift terms on the atomic inversion dynamics when a multilevel
atom interacts with a two-mode quantized radiation field have been
also investigated by Cardimona et al. \cite{Cardimona2}. They have shown
that these Stark terms are relatively unimportant for Raman scattering
but could not be ignored in equal- or unequal-frequency
absorption.

Very recently a class of coupled-channel cavity QED models
describing two-photon resonant Raman processes in multiple-
$\Lambda$ configuration has been solved analytically by an algebraic
method based on the introduction of an angular momentum algebra for this
multiwave mixing system \cite{Wu}.

\subsubsection{Nondegenerate two-mode two-photon JC model }

The system we consider in this subsection is a generalized JC
model in which the atomic transitions are mediated by
nondegenerate two-photon absorption or emission. As it is well
known, in the two-photon processes at least an intermediate level,
assumed to be coupled to the ground and excited states
respectively by dipole allowed transitions, is involved. As shown
in Fig.~\ref{fig4} we denote by $E_g$, $E_i$ and $E_e$ the lower,
intermediate and upper atomic energy levels and, indicating by
$\omega_1$ and $\omega_2$ the frequencies of the two cavity modes,
suppose that the conditions $E_i-E_g=\omega_1-\Delta$,
$E_e-E_i=\omega_2+\Delta$ are satisfied. It is possible to demonstrate
that, if $\vert \Delta \vert$ is much larger than the one-photon
Rabi frequency of the oscillations between $\vert i \rangle$  and
$\vert g \rangle$ and between $\vert i \rangle$ and $\vert e
\rangle$ respectively, the effective Hamiltonian model describing
the system under scrutiny assumes the form:
\begin{equation}\label{hnd}
  \hat{H}_{nd}=\omega_1\hat{a}_1^{\dag}\hat{a}_1
  +\omega_2\hat{a}_2^{\dag}\hat{a}_2+(\omega_0+\beta_2\hat{a}_2^{\dag}\hat{a}_2-\beta_1\hat{a}_1^{\dag}\hat{a}_1)\hat{S}_z+
(\lambda \hat{a}_1\hat{a}_2\hat{S}_++h.c.)
\end{equation}

The alkali-metal Rydberg energy diagram, for example, provides
concrete examples fulfilling the conditions leading to an atom -
bimodal cavity interaction Hamiltonian such as $\hat{H}_{nd}$
\cite{Brune87,Maia-Nieto}. Observe that the hamiltonian model
given by equation \ (\ref{hnd}), contains an
intensity dependent detuning described by the two terms
proportional to $\beta_1$ and $\beta_2$ respectively, whereas the
constant $\lambda$ measures the effective two-photon atom-field
coupling. This Hamiltonian  is exactly solvable and a unitary
operator accomplishing the canonical dressing of the two-level
atom by the bimodal cavity field, has also been constructed
\cite{NMdressed}.

The Hamiltonian model $H_{nd}$ was
firstly introduced by Gou \cite{Gou1,Gou2}.
In \cite{Gou1}, neglecting the Stark shift
arising from the presence of virtual levels, that is putting in
equation \ (\ref{hnd}) $\beta_1\sim \beta_2\sim 0$, the quantum
dynamics of the atom and the temporal evolution of the field
statistics were studied. In particular the author considers two
types of initial state of the field, namely ordinary uncorrelated
coherent states in both modes and the two-mode squeezed vacuum
state where the number of photons in one of the modes is equal to
the number of photons in the other one. It has been noted that the
dynamics of the model is qualitatively different in correspondence
of these two kinds of states.  In more detail it has been shown
that new features of quantum collapse and revival occur in
presence of the two-mode squeezed vacuum. The revivals here appear
periodically sharp.
The origin of such a behaviour is
strictly related to the particular class of Rabi frequencies
involved in the dynamics of the respective physical systems.
If, on the contrary, the field is initially prepared in a product
of two coherent states, the atom exhibits the phenomena of
collapse and revival as well, but no periodicity can be observed.
The different behaviour of the atomic dynamics in correspondence
to initial correlated and uncorrelated cavity field has also been
put into evidence by Joshi and Puri \cite{J-P} as well as by Gerry
and Welch \cite{G-W}. In both papers the Stark shift terms are
neglected.

In  \cite{Ashraf}  Ashraf has investigated the cavity field
spectra for various initial fields of a nondegenerate two-photon
JC model neglecting the two terms proportional to $\beta_1$ and
$\beta_2$ respectively.

The results presented in \cite{Gou1,G-W,Ashraf}
 could be qualitatively or quantitatively affected by the
presence of the Stark shift of the two effective two atomic
levels. As shown by Gou \cite{Gou2}, indeed, in most circumstances
these terms cannot be ignored, even if for certain correlated
two-mode field states, the elimination of Stark shifts is possible
at least when one evaluates the ensemble averages relevant to the
phase insensitive terms. However, for the phase sensitive averages
such as the squeezing of cavity fields, the Stark shifts still
cannot be ignored in any case. Also Cardimona et al
\cite{Cardimona2}  have proved that, differently from the Raman
coupled model, for unequal-frequency absorption or emission, the
Stark effect could not be negligible.

\subsubsection{Intensity dependent two-mode two-photon JC model} In
this section we investigate on the nondegenerate two-mode
two-photon JC hamiltonian model assuming that both the coupling
strenght and the atom-field detuning may depend on the populations
of the two modes. The hamiltonian  describing such a general
nonlinear model may be written down as \cite{NMdressed}:
\begin{equation}\label{hdressed}
\hat{H}_{id}=\hat{H}_1+\hat{H}_2+\hat{H}_3
\end{equation}
with
\begin{equation}\label{hdressed123}
\hat{H}_1=\omega_1\hat{a}_1^{\dag}\hat{a}_1+\omega_2\hat{a}_2^{\dag}\hat{a}_2; \;\;\;\;\;
\hat{H}_2=\omega_0G(\hat{a}_1^{\dag}\hat{a}_1,\hat{a}_2^{\dag}\hat{a}_2); \;\;\;\;\;
\hat{H}_3=\lambda[\hat{a}_1\hat{a}_2F(\hat{a}_1^{\dag}\hat{a}_1,\hat{a}_2^{\dag}\hat{a}_2)\hat{S}_++h.c.]
\end{equation}
where the RWA has been used. The two functions
$G(\hat{a}_1^{\dag}\hat{a}_1,\hat{a}_2^{\dag}\hat{a}_2)$ and
$F(\hat{a}_1^{\dag}\hat{a}_1,\hat{a}_2^{\dag}\hat{a}_2)$
appearing in equation \ (\ref{hdressed123}), are introduced to take into
account the possibility of intensity dependence both in the
atom-field detuning and in the coupling constant.
Let's observe that putting in particular $\omega_0G(\{\hat{a}_i^{\dag}\hat{a}_i\})=\omega_0-\beta_1\hat{a}_1^{\dag}\hat{a}_1+\beta_2\hat{a}_2^{\dag}\hat{a}_2$ and $F(\{\hat{a}_i^{\dag}\hat{a}_i\})=I$ we recover the nondegenerate two-mode two-photon JC model discussed in the previous section.
In  \cite{NMdressed} a symmetry-based detailed construction of the
atom-cavity eigenstates as well as of the energy spectrum of the
dressed atom is presented. The starting point of such a
construction consists in finding canonical symmetry
transformations for $\hat{H}_{id}$ and the related constants of
motion.
In particular they prove that the operators
$\hat{N}=\hat{a}_1^{\dag}\hat{a}_1+\hat{a}_2^{\dag}\hat{a}_2+2\hat{S}_z+1$ and $\hat{D}=\hat{a}_1^{\dag}\hat{a}_1-\hat{a}_2^{\dag}\hat{a}_2$ are constants of motion indipendently from the specific form of the two intensity dependent functions $F$ and $G$.
The existence of these two constants of motion turns out to be
very useful in the construction of a unitary operator
diagonalizing $\hat{H}_{id}$. It is worth noting that the
symmetry-based conclusions envisaged in  \cite{NMdressed} may be
applied whatever  the analytic dependence of the
intensity-dependent terms on the populations of the two-cavity
modes is. This result allows to highlight the role of the
nonlinear contributions on the two-level atom dressed by the
cavity field. In order to do this, in \cite{NMdressed} the authors
study the quantum dynamics of the initially excited two-level atom
interacting, in the context of the general model, with a cavity
field prepared in the tensorial product of a coherent state and a
Fock state. Exploiting the flexibility of their results the
authors compare the atomic evolution of several particular models
finding a rich dynamical behaviour dominated by the presence of
collapses and revivals of the Rabi oscillations. The occurrence of
these collapses and revivals is a common phenomenon to all the
models considered in \cite{NMdressed} even if some other
distinctive features characterize the time evolution of the atomic
inversion in each individual model.

The existence of sensitivity to the population of the field mode
initially prepared in a Fock state is also pointed out. This
effect is of quantum nature and it is particularly noteworthy
since, as showed in the subsequent subsection II G 2 it leads to
proposals for generating number states in a single cavity mode
starting from the vacuum.

\subsubsection{Degenerate two-mode two-photon JC model. Parity Effect }

Suppose now that the two modes of the cavity field have the same
frequency $\omega$ assuming that they differ either by the
polarization vectors or by the direction of propagation. The
degeneration condition imposed to the two quantized
electromagnetic modes makes the effective hamiltonian model
describing the two-photon interaction between the bimodal cavity
and a single few level atom, very different from that adopted in
the previous section. The
effective model obtained in this degenerate case is not a trivial
generalization of the nondegenerate one \ (\ref{hnd}). The
explicit construction of the effective Hamiltonian  for the
degenerate quadratic atom-field interaction leads to the following model:
\begin{equation}\label{hd}
\hat{H}_{d}=\hat{H}_0^d+\hat{H}_R^d+\hat{H}_S^d+\hat{H}_I^d
\end{equation}
with
\begin{equation}\label{hd0rs}
\hat{H}_0^d=\omega_0 \hat{S}_z+\omega \sum_{\mu=1,2}\hat{a}_{\mu}^{\dag}\hat{a}_{\mu};  \;\;\;\;
\hat{H}_R^d=(r_1\hat{a}_1\hat{a}_2^{\dag}+r_2 \hat{a}_1\hat{a}_2^{\dag}\hat{S}_z)+h.c.; \;\;\;\;
\hat{H}_S^d=s \hat{S}_z \sum_{\mu=1,2}\hat{a}_{\mu}^{\dag}\hat{a}_{\mu}
\end{equation}
and
\begin{equation}\label{hdi}
\hat{H}_I^d=(\sum_{\mu=1,2}\lambda_{\mu}\hat{a}_{\mu}^2+g\hat{a}_1\hat{a}_2)\hat{S}_+ +h.c.
\end{equation}
This Hamiltonian model has been recently introduced by A. Napoli
and A. Messina \cite{NM1997}. The structure of $\hat{H}_d$
reflects the fact that the atom-mode-mode field energy exchanges
in the degenerate case may take place following all the different
possible channels compatible with the conservation of the total
excitation number operator
$\hat{N}=\hat{a}_1^{\dag}\hat{a}_1+\hat{a}_2^{\dag}\hat{a}_2+2\hat{S}_z+1$.
It is easy,indeed, to
demonstrate that the operator $\hat{N}$ commutes with $\hat{H}_d$
whatever the coupling constant $s$, $r_1$, $r_2$, $\lambda_1$,
$\lambda_2$ and $g$ are. The term of $\hat{H}_d$  depending on the
constant $g$, for example, arises from the possibility that the
atom makes transitions involving a simultaneous gain or loss of
one photon in both modes. The contributions related to the
coupling constants $r_1$ and $r_2$  describe, on the other hand,
mode-mode energy exchanges not involving the atom or changes in
its state. These terms are known as Rayleigh terms. Stark terms
describing a field-dependent detuning appear in $\hat{H}_d$  with
the coupling constant $s$. Finally the possibility of two-photon
exchanges between the atom and each cavity mode is described by
the terms of $\hat{H}_d$ containing $\hat{a}_{\mu}^2\hat{S}_+$.
The strength of this specific effective atom-field interaction
mechanism is measured by $\lambda_{\mu}$.

In what follows we concentrate our attention on the case
correspondent to $g=0$ and $\lambda_1=-\lambda_2=\lambda$, that is
on all the physical systems effectively describable by means of
the following Hamiltonian model:
\begin{eqnarray}\label{hdsempl}
\tilde{H}_d&=&\omega_0 \hat{S}_z+\omega
\sum_{\mu=1,2}\hat{a}_{\mu}^{\dag}\hat{a}_{\mu}+s \hat{S}_z
\sum_{\mu=1,2}\hat{a}_{\mu}^{\dag}\hat{a}_{\mu}+ [(r_1\hat{a}_1\hat{a}_2^{\dag}+r_2
\hat{a}_1\hat{a}_2^{\dag}\hat{S}_z)+h.c.]\\ \nonumber
&+&[\lambda(\hat{a}_1^2-\hat{a}_2^2)\hat{\hat{a}S}_+
+h.c.]
\end{eqnarray}
In  \cite{NM1997} it has been shown that specific physical
situations in correspondence of which the conditions $g=0$ and
$\lambda_1=-\lambda_2=\lambda$ are satisfied, may be envisaged.

It is possible to convince oneself that in this case the Hamiltonian $\tilde{H}_d$ possesses another hermitian
constant of motion, namely
$\hat{C}=\hat{a}_1^{\dag}\hat{a}_2+\hat{a}_1\hat{a}_2^{\dag}$. The
existence of the two constants of motion, $\hat{N}$ and $\hat{C}$,
is at the basis of the possibility of exactly evaluating the
dynamics of the system bringing to  light the existence of an
intrinsically quantum phenomenon, christened in \cite{NM1997} \lq
\lq parity effect\rq \rq, strictly connected with the granularity
of the radiation field. In order to describe this nonclassical
effect, let's suppose to inject a two-level atom, initially
prepared in its ground state, into the cavity where one mode is
excited in a state exactly containing $n$ photons whereas the
other one is left in its vacuum state. After an appropriate
$n-$dependent interval of time, the system described by
$\tilde{H}_d$ exhibits a peculiar sensitivity to the parity of the
initial total number of photons $n$. Stated another way, the
physical system distinguishes between the two different initial
conditions $n$ odd or $n$ even, showing macroscopically different
quantum behaviours. This sensitivity to the parity of the initial
photon number clearly manifests itself in the dynamics of the mean
value of the population operators
$\hat{a}_{\mu}^{\dag}\hat{a}_{\mu}$. In Fig. ~\ref{fig5} we in
particular display the time evolution of the $\mu =1$ mode average
photon number, correspondent to the initial conditions  $n=20$
(Fig.~\ref{fig5}(a)) and $n=21$ (Fig.~\ref{fig5}(b)) respectively,
putting $r_1=r_2=s=0$. These plots show that the initial behaviour
of the temporal variation of the photon number in the mode $\mu
=1$ is characterized by a rapid reduction to a half and it is
followed by a relatively longer interval of time dominated by a
negligible photon exchange between the two modes. After such a \lq
\lq lethargy period\rq \rq, the time evolution of the mode
population appears to be very sensitive to the parity of the
initial photon number. Indeed, choosing the initial condition
$n=20$ the field mode completes the transfer of its population to
the other mode provoking an inversion in the dynamical role
between the two modes. On the other hand, the plot of the mean
value $\bar{n}_1(t)$ of the operator
$\hat{n}_1=\hat{a}_1^{\dag}\hat{a}_1$, relative to the initial
condition $n=21$, shows that, after the \lq \lq lethargy period\rq
\rq , the field mode reabsorbs the photons already transferred to
the initially empty mode.

The existence of this interesting nonclassical effect, that is the
sensitivity to the parity of the initial number of photons present
in the excited field mode, is a peculiar feature of the
atom-degenerate bimodal cavity field interaction described by
$\tilde{H}_d$ and, as demonstrated in \cite{NM1997}, it does not appear related to the simplifying
assumption $r_1=r_2=s=0$.

We wish to underline that, as we will show
in the next section, the parity effect can be successfully
exploited for generating highly nonclassical bimodal field states
having the form of Schr\"{o}dinger cat states
\cite{NMmysteries,NMcrrnsm}.

Before concluding we wish to anticipate that in the third section
of this paper, the parity effect shall appear also in the dynamics
of some hamiltonian models of interest in trapped ion context.
Such an occurrence is by no means surprising in view of the fact
that, at a formal level, CQED and trapped ion domains make use of
similar models. The important point to stress is however that, due
to the different meaning attributed to the same hamiltonian model
within the two experimental realms, very different dynamical
consequences of the parity effect in these two context, shall be
brought to light.

\subsection{Generation of nonclassical field states in a bimodal
high-$Q$ microcavity}

Investigating the dynamical behaviour of a physical system from a
theoretical point of view, very often involves more or less tacit
assumptions about the possibility of preparing the system in a
given state at $t=0$. In this sense the experimental realization
of assigned initial conditions for a system  is  a prerequisite
for studying fundamental aspects of quantum theory \cite{Haroche1,Braginski}. Besides this
inherent theoretical interest, the state preparation plays a
central role also in the context of the new field of Quantum
Information theory encompassing Quantum Computing and Quantum
Cryptography \cite{Ekert,Zeilinger,Bennet1,Ekert1}.

So far two approaches have been mainly exploited in order to
produce desired prefixed states. The first one consists in finding
the appropriate Hamiltonian which can transform a given initial
state into a desired one via unitary evolution. The other commonly
used approach consists in making a quantum measurement on one part
of an entangled system provoking the wave function collapse of the
other subsystem.

Over the past few years, also under the stimulus of this rapid
development in the experimental techniques, several ingenious
procedures for generating nonclassical states of the
electromagnetic field have been reported in the realm of cavity
quantum electrodynamics
\cite{Rempe,Krause,Garraway,Gerry,Kurizki,Davidovich,NM-I,NM-II,Haroche98,Napoli,Weidinger,Englert,Haroche01}.
Particular attention has been devoted to the creation of
single-mode number states in view of the fact that they are
considered as prototype of nonclassical state
\cite{Krause,Garraway,Kurizki,NM-I,NM-II,Napoli,Weidinger}. These
states, corresponding to an  exact number of photons, possess in
addition maximum information capacity and therefore they might be
used for encoding and processing quantum information.

A lot of attention has moreover been
devoted to the possibility of preparing superpositions of two or more
macroscopically distinguishable states \cite{NMmysteries,NMcrrnsm,Garraway,Haroche98,Brune96}. Such  states, called Schrodinger cat
like states, provide an useful starting point to progress with the
understanding of decoherence phenomena and to gain more insight about the
elusive border between classical and quantum worlds \cite{Zurek}.

Most of the proposals present in literature deal with the
generation of nonclassical states of a single cavity mode
\cite{Rempe,Krause,Garraway,Kurizki,Davidovich,NM-II}. The
interesting dynamical properties manifested by the two-mode models
discussed in the previous sections, suggest that also the
generation of two-mode field states might play an important role
in highlighting fundamental aspects of quantum theory. In the
present section we will show that the interaction between single
atoms and two independent modes of a bimodal cavity can be
successfully exploited to generate several nonclassical states of
the bimodal field. At the same time, resuming an idea already
present in the paper of Gou \cite{Gou1}, we will demonstrate that
the bimodal cavities can be also used for producing, with
reasonable simple protocols, nonclassical states of a single mode
of the cavity. In particular, in subsection II G 2 we will prove
that when the attention is focused on a single radiation mode, the
presence of the other one plays the relevant role of a \lq \lq
control subsystem\rq \rq.

\subsubsection{Preparation of bimodal field states}

The basic idea of most of the experimental procedures aimed at
generating assigned nonclassical states of one or more modes of a
single cavity field, can be traced back to the possibility of
manipulating the statistical properties of the radiation field by
injecting, one at time, a flux of Rydberg atoms into the cavity.

In the framework of micromaser-based preparation of cavity field
states, one of the methods generally used for guiding the field
toward a prefixed state is the so-called conditional measurement
approach which consists in selecting only those sequences wherein
each atom exits the cavity in a given state and discarding all
others \cite{Garraway,Gerry,Kurizki}. Although most of the
experimental schemes exploiting this idea are conceptually very
attractive, nevertheless they are often not mature enough to be
qualified as realistic projects. Generally speaking, indeed, the
high degree of precision assumed for the experimental apparatus as
well as the fact that, even in such idealized conditions, the
probability of success of the experiment is often very low,
seriously counter the possibility of practically implementing
these proposals.

\vspace{1cm}

{\it a)   Correlated bimodal states from atomic coherences}

Consider, for example, the controllable generation procedure aimed
at producing a class of arbitrary two-mode field states in a
cavity, recently proposed by Deb et al \cite{Deb2}. Their proposal
is based on the possibility of transferring atomic coherences to
the cavity field exploiting the Raman coupled model.
The authors  show that,
injecting into the cavity $N$ three-level atoms each one initially
prepared in a suitable superposition of its two nondegenerate
ground states and measuring their internal state after the
interaction with the cavity field, it is in principle possible to
guide the field toward a state $\vert \phi
\rangle=\sum_{n=0}^{N}\pi_n^{N} \vert n,N-n \rangle$ starting from
$\vert \psi(0)\rangle=\vert 0,N \rangle$. Of course $\vert \phi \rangle$ is normalized and its probability amplitudes $\pi_n^{N}$ is determined by the coupling mechanism and by details of the protocol.

The idea firstly exploited by Vogel et al. \cite{Vogel93} of transferring
atomic coherences to the field, has also been  used by Zheng
and Guo \cite{Guo1} who propose a rather formal way of preparing multimode field states having equal photon numbers in these modes.

The nonresonant interaction of a two-level atom and a bimodal cavity field is exploited by Guo and Zheng \cite{Guo2} to propose a
method for generating, at least in principle, any linear
superposition of product of coherent states of the bimodal field.
The basic idea of their scheme is that the photon statistical
distribution of the field will not be changed by the
interaction if the detuning between
the atomic transition frequency and the frequency of each mode is
much larger than the coupling constant.
On the contrary the phase of the field evolves with the time.
 The authors claim that, by state-selective measurements on the atom,
it is possible to produce entangled coherent states of the bimodal
field provided that both the two modes are initially prepared in a
coherent state.

\vspace{1cm}

{\it b) Pair Fock states}

 In  \cite{NM-I} the theory underlying a new experimental
 conditional measurement scheme for the preparation of selected
 equal-intensity bimodal Fock states is presented,  carefully
 discussing its practical feasibility.  The existence of a reliable
 experimental method to produce pair-Fock
 states of controllable intensity, besides its inherent theoretical
 interest, is also of relevance from an applicative point of view.
 In the field of quantum communications, in fact, it has been very recently analyzed the possibility of transmitting, with relatively high
 capacity, information by multiple parallel \lq \lq number state
 single channels\rq \rq \cite{Caves}. The experimental project presented
 in \cite{NM-I} is based on the nondegenerate two-mode two-photon
 JC coupling, described by equation \ (\ref{hnd}),
 between single Rydberg atoms and a bimodal cavity.
 A peculiar and noteworthy characteristic of this specific
 atom-two mode field interaction mechanism
 is the possibility of obtaining, starting from an empty cavity,
 pair-Fock states of the bimodal field with a relatively
 high intensity.
The building up of an equal-intensity pair-Fock state of the
microresonator field, in accordance with the proposal under
scrutiny, results from the success of a multiple iterative event.
Suppose the cavity prepared in its vacuum state  $\vert 0,0
\rangle$. A single excited atom enters the cavity where it
simultaneously exchanges one photon with each mode as effectively
described by  equation \ (\ref{hnd}).
Immediately after the atom
has left the cavity its internal state is measured using an
appropriate detection technique. It is possible to convince
oneself that an atomic ground state outcome disentangles the atom
from the field which, in turn, collapses in the state $\vert 1,1
\rangle$. At this point a second atom is injected into the cavity
with an appropriate velocity. If even the second atom is detected
in its ground state exiting the cavity, then the field in the
microresonator is projected onto its Fock state  $\vert 2,2
\rangle$. It is not difficult to convince oneself that this
procedure may be iterated so that a successful sequence of $N$
atoms detected in their ground states immediately after they have
left the cavity, builds up the equal-intensity pair-Fock state
$\vert N,N \rangle$ of the bimodal field. The authors present a
detailed  quantitative analysis of the scheme showing, indeed,
that its probability of success is of practical interest also in
correspondence to high intensity pair Fock states $(N\sim 20)$. On
the other hand, notwithstanding the lowering of such a probability
due to the presence of important technological limits of the
experimental apparatus, the scheme maintains a reasonable
compatibility with the presence of sources of imprecision in the
concrete realization of the experiment.

A common aspect to many micromaser-based proposals for
generating nonclassical states of the cavity field is that they
require the availability of ideal devices with which the
interaction time between each atom and the field may be sharply
selected.
In connection with the generation of bimodal Fock states,  the
presence of unavoidable fluctuations in the atom-cavity
interaction times is taken into account from the beginning in
\cite{Napoli}. The theory developed in this paper clearly shows
that the impossibility of sharply controlling the interaction time
between each atom and the bimodal cavity, as well as the presence
of other important sources of experimental imprecision, reduce the
probability of success for the preparation of equal-intensity pair
Fock states to values that, however, are still of experimental
interest. As stressed in \cite{Napoli}, this circumstance is
strictly related to the interaction mechanism chosen for
manipulating the cavity field.

\vspace{1cm}

{\it  c) Maximally entangled states}

The nondegenerate two-mode two-photon JC coupling given by equation \ (\ref{hnd}) appears to be a good basis for generating other
classes of purely quantum states as, for example, maximally
entangled two-mode states having the form $\vert
\phi_{\pm}\rangle=\frac{1}{\sqrt{2}}[\vert 0,0 \rangle {\pm}
e^{i\phi}\vert 1,1 \rangle] $ \cite{NMjmo2000}.
An attractive and relevant aspect of the scheme
reported in \cite{NMjmo2000}, is that the phase factor $\phi$ can
be manipulated from $-\pi$ to $\pi$ by simply varying an
experimentally easily controllable parameter.

Very recently, the possibility of entangling two independent
cavity modes has been experimentally demonstrated
\cite{Haroche01}. In particular, in  \cite{Haroche01} the authors
show that, using a single circular Rydberg atom,  two quantized
modes of a superconducting cavity can be prepared in the following
maximally entangled state characterized by the fact that the two
modes share a single photon:
\begin{equation}
\frac{1}{\sqrt{2}}(e^{i \phi}\vert 0,1 \rangle+\vert 1,0 \rangle)
\end{equation}
This EPR pair state is then revealed by a second atom probing, after a delay, the correlations between the two cavity modes.

\vspace{1cm}

{\it d)  Photon number statistics invariant manipulation}

Exploiting the same interaction mechanism envisaged in the
previous section, it is also possible to modify only the phases of
the initial probability amplitudes of finding a well-defined
population in the bimodal field simply injecting a single
two-level Rydberg atom into the cavity \cite{NMoc}. As shown in
\cite{NMoc}, this kind of radiation field manipulation can be
useful, for example, in the construction of states orthogonal or
quasi-orthogonal to the initial one or for controlling the
squeezing parameter of a given single-mode state.

\vspace{1cm}

{\it e)  Schr\"{o}dinger cat states from parity effect}

The existence of the intrinsically nonclassical \lq \lq parity
effect\rq \rq discussed in section II F 4, is at basis of the
possibility of generating entangled bimodal states having the
nature of Schr\"{o}dinger cats \cite{NMmysteries,NMcrrnsm}. We
wish to underline that in this case the two modes of the bimodal
cavity are degenerate and the hamiltonian model describing the
two-photon interaction between the field and a single two-level
Rydberg atom is given by equation \ (\ref {hdsempl}).
Suppose to prepare  the cavity leaving one mode in its vacuum
state and exciting the other one in a coherent state $\vert \alpha
\rangle$. Such an initial condition can be easily realized in
laboratory. The vacuum state is, indeed, reached controlling the
temperature at which the experiment is performed \cite{varie}. On
the other hand, it is possible to excite one mode in a coherent
state $\vert \alpha \rangle$, having a prefixed intensity, using
classical currents \cite{Haroche1}. The experimental protocol
described in  \cite{NMcrrnsm}, starts injecting in the resonator
the two-level atom prepared in its ground state. In order to
elucidate in which way the parity effect previously recalled may
be exploited for reaching the prefixed scope, it is convenient to
cast the initial condition imposed to the system as
 sum of the following two contributions:
\begin{equation}\label{SC0}
\vert \psi(0) \rangle=e^{- \frac{\vert \alpha \vert^2}{2}}
\sum_{n=0}^{\infty} {\frac{\alpha^{2n}}{\sqrt{(2n)!}}\vert 2n,0,-
\rangle}+e^{- \frac{\vert \alpha \vert^2}{2}}
\sum_{n=0}^{\infty}{\frac{\alpha^{2n+1}}{\sqrt{(2n+1)!}}\vert
2n+1,0,- \rangle}
\end{equation}

In view of the linear
character of quantum mechanics and as a consequence of the parity
effect, the initial state evolves toward two macroscopically
different states respectively stemming from the two orthogonal
contributing terms appearing in equation\ (\ref{SC0}). In
particular, it is possible to analytically evaluate the instant of
time, $t_{\vert \alpha \vert}$, at which the system exhibits in
the best way such parity effect-dependent consequences. Thus,
manipulating the atomic velocity in such a way that the
interaction time between the cavity field and the two-level
Rydberg atom exactly coincides with  $t_{\vert \alpha \vert}$ and
measuring the internal atomic state immediately after the atom
leaves the cavity, the bimodal field is projected into a state
$\vert \psi \rangle$ having the desired character of a bimodal
Schr\"{o}dinger cat. Indeed the state $\vert \psi \rangle$ can be
written as a superposition of two orthogonal field states, $\vert
\varphi \rangle$ and $\vert \chi \rangle$, macroscopically
distinguishable in the sense that in the bimodal state $\vert
\varphi \rangle$ the field energy is essentially concentrated in
the initially empty mode whereas in the bimodal state $\vert \chi
\rangle$  all the energy of the field may be found in the other
mode.

\subsubsection{Using bimodal cavities for generating single-mode
nonclassical states}

The interaction between single Rydberg atoms and bimodal high-$Q$
cavities can be also exploited with success for generating
interesting nonclassical states of a single cavity mode. As we
will show in this section, the presence of a second quantized mode
of the electromagnetic field turns out to be very useful playing the
role of an auxiliary mode whose initial condition may be
appropriately chosen in order to optimize single-mode nonclassical
state generation procedures.

In  \cite{NM-II} for example, a bimodal high-$Q$ cavity is used
for the controlled preparation of single-mode Fock states.

This method, developed in the context of the
nondegenerate two-photon micromaser, is based on appropriate
sequences of wave packet reductions so that its degree of success
is accordingly measured in terms of a conditional probability.
Qualitatively speaking, the main result reached in this article is
the possibility of conjugating experimentally acceptable values of
such a probability with a reasonable high value of the number of
photons deposited in the cavity mode of interest. Following the
method presented in  \cite{NM-II} the generation, for example, of
the Fock state $\vert 10 \rangle$ is greater than 0.1. The
practical reliability of the proposal reported in \cite{NM-II} is
moreover carefully discussed from several points of view. In
particular the nonideal behaviour of the atomic velocity selectors
today available, has been taken into account by the authors. Their
analysis puts clearly into evidence that the procedure is
characterized by an experimentally significant stability against
the unavoidable uncertainty in the atom-field interaction times.

In the context of this topical review, it is important to point
out that, as shown in \cite{NM-II}, the possibility of having
experimentally interesting values of the probability of success,
directly stems both from the peculiar two-photon two-mode
interaction mechanism adopted in the paper, and, above all, from
the existence of an auxiliary mode. In the context of their
proposal, in fact, the initial photon distribution of such a mode
is singled out looking for the best compromise between purely
mathematical optimization arguments, concerning the probability of
success of the scheme itself, and practical limitations stemming
from considerations of an experimental nature. Guided by these
considerations, the bimodal cavity is prepared leaving the mode of
interest in its vacuum state and exciting the auxiliary one in a
coherent state $\vert \alpha \rangle$ with intensity $\vert \alpha
\vert ^2$ appropriately chosen.
Starting from this initial condition and inverting the role of the
two modes, that is considering as mode of interest the one
prepared in the coherent state, it is also possible to generate a
class of quantum superpositions of two single-mode coherent states
each one having the same intensity of the initial state
\cite{NMqs00,NMacta99}.  In more detail, exploiting the passage of
one atom only through the bimodal cavity initially prepared in
accordance with the request previously discussed, a conditional
measurement of the internal atomic state, after the atom has left
the resonator, projects the mode of interest in a state having the
following form:
\begin{equation}\label{coher}
\vert \Psi \rangle=A[\vert \beta \rangle -e^{-i\lambda\tau}\vert
\beta e^{-i\lambda\tau}\rangle]
\end{equation}
$A$ being an appropriate normalization constant.  In equation\
(\ref{coher}) we have indicated respectively by $ \lambda $ and $
\tau $ the coupling constant and the interaction time between the
atom and the bimodal cavity. Moreover $\beta=\alpha
e^{-i\gamma_-(\tau)} $ and
$\gamma_-(\tau)=(\frac{\lambda}{2}-\omega_1)\tau$,   $\omega_1$
being the frequency of the mode prepared in the state \
(\ref{coher}).
In the next section we present a class of dynamical problems
studied in the trapped ion domain. This class includes some of the
problems examined in the course of this section reinterpreted in
the new experimental context. In addition we shall study the
dynamical properties of other hamiltonian models extraneous to
CQED domain, describing the laser driven coupling of internal and
external ionic degrees of freedom, still in terms of pseudo-spin
and two bosonic dynamical variables.

%%%%%%%%%%%%%%%%%%%%%%IONI IN TRAPPOLA %%%%%%%%%%%%%%%%%%%%%%%%%%%%%%%%%%%%
%%%%%%%%%%%%%%%%%%%%%%%%%%%%%%%%%%%%%%%%%%%%%%%%%%%%%%%%%%%%%%%%%%%

\pagebreak

\section{Ions in bidimensional traps}

\subsection{Introduction} Recent progress in laser cooling and
trapping techniques has opened up new perspectives for research in
the fields of atomic physics and quantum optics. By cooling down
ions to very low temperatures, indeed, purely quantum
manifestations become observable \cite{lcooling,cooltraprev}.

A single ion confined in a Paul microtrap is theoretically
equivalent to a particle, moving in a harmonic potential, whose
center of mass (c.m.) motion is quantized as harmonic oscillator.
By irradiating the ion with classical laser beams, its internal
(electronic) and motional degrees of freedom can be coupled due to
the momentum exchange between the laser field and the ion.

It has been shown \cite{blockley,cirac1,nist1,blatt1} that there
exist suitable conditions under which the effective Hamiltonian
describing such a system has a form similar to the JCM where the
quantized radiation field is replaced by the quantized ionic c.m.
motion. This circumstance suggested the idea of studying phenomena
typical of the cavity QED context with trapped ion systems. A
comparison of experiments measuring quantum Rabi oscillations in
the  cavity QED and trapped ion contexts \cite{nist1,harRabi}
shows that the coherence time characterizing the dynamics of
single trapped ions is at least one order of magnitude bigger than
that typical of the behaviour of atoms crossing high Q cavities.
An important aspect which makes the trapped ion systems so
attractive and interesting is, indeed, the fact that they are very
well isolated from external environment being thus  very good
candidates for experiments relying on the persistence of quantum
coherences. Moreover, by changing the lasers configuration and
parameters, a huge class of effective Hamiltonians may be
engineered. These facts, together with the possibility of
measuring the internal ionic state with almost unit efficiency,
gave rise to many proposals extending Feynman's ideas on the use
of a given quantum system to simulate the behaviour of another
quantum system \cite{feynman}. As far the topic of this paper is
concerned, many bimodal interactions and phenomena typical of both
cavity QED and nonlinear optics, can be easily obtained in the
trapped ion context. For example the vibrational analogue of
parametric amplification, Kerr-type nonlinearities, multimode
mixing and multiphoton down conversion can be realized
\cite{vog1,vog2,agarw1}. Moreover a Mach-Zender interferometer
acting on two modes of oscillation of the trapped ion may be
simulated allowing the realization of entangled states of the
ionic vibrational 2D motion \cite{MZinterfSU(2),bibbiaNIST,nist2}.
Also maximally entangled states and Bell states may be engineered
by means of several different kinds of easily realizable bimodal
interactions\cite{bibbiaNIST,nist2}.

It is worth noting, however, that when the spatial wave function
representing the c.m. motion is no longer small compared with the
driving laser wavelength, there appear nonlinear effects,
characteristic of the trapped ion system, which have no
counterpart in standard nonlinear optics \cite{matos1,matos2}.
Under these conditions interesting phenomena appear due to the
overlap of the atomic center of mass wave function with the laser
waves causing, nonlinear modification of the coupling strength
\cite{vog3,vog1,vog2}.

Summing up, trapped ion systems, on one hand allow to verify
experimentally many interesting phenomena which are difficult to
observe in other systems, and on the other hand they present a
very rich quantum dynamics showing nonclassical effects having no
analogue in optical systems. The considerations we have done
justify the growing interest toward such systems. In what follows
we will present a review of the studies on bidimensionally trapped
ions.

We first describe briefly the basic principles of the Paul
microtrap and of the techniques for detecting the ionic internal
states. The effective Hamiltonian describing the ionic dynamics is
then derived and some proposals for engineering  generic bimodal
interactions  are briefly reviewed. In subsection III B we
concentrate our attention on the dynamics of the laser driven
system and, in particular, on the appearance of some nonclassical
effects in the ionic dynamics. Finally, in subsection III C
several methods for generating different types of  nonclassical
states of the ion are discussed.

\subsubsection{Quantum motion of ions in Paul traps} Due to their
net charge, atomic ions can be confined by particular arrangements
of electromagnetic fields. To study ions at low energy, two types
of traps are usually used: the Penning trap, which uses a
combination of static electric and magnetic fields, and the Paul
or rf trap, which confines ions primarily through ponderomotive
forces generated by inhomogeneous oscillating fields. The
operation of these traps is discussed in various reviews
\cite{cooltraprev,trapprev} and in a recent book by Ghosh
\cite{ghosh}.

Over the last few years many experimental results for ions
confined in Paul microtraps have been presented. For this reason
we choose to concentrate our review to physical situations in
which the trapping apparatus is a Paul microtrap. In
Fig.~\ref{fig:1} we show a schematic diagram of such a trap. If a
constant voltage $U>0$ and an rf alternating voltage $V \cos
(\omega_{rf}t)$ are applied to the trap electrodes, an ion of
charge $q$ and mass $m$ moving inside the trap sees a time
dependent potential of the form

\begin{equation}
\hat{\Phi} (\hat{x},\hat{y},\hat{z},t)= q \left( Q_{xx}(t)
\hat{x}^2 + Q_{yy}(t) \hat{y}^2 + Q_{zz}(t) \hat{z}^2 \right)
\label{1}
\end{equation}
where $x, y$ and $z$ are the coordinates of the center of mass
motion of the ion and $Q_{ii}(t)$ ($i=x,y,z$) are the diagonal
elements of the traceless tensor of the quadrupole moment, given
as

\begin{equation}
Q_{ii}(t) = Q_{ii} \left[ U + V \cos (\omega_{rf}t)\right]
\hspace{1cm} (i=x,y,z) \label{2}
\end{equation}
From equation (\ref{1}) and (\ref{2}) the equations of motion of
the ionic c.m. coordinates can be derived in the form of Mathieu
differential equations. Stable solutions of these equations are
superpositions of a c.m. oscillation   at the driving frequency
$\omega_{rf}$, usually called micromotion, and oscillations of
typically much smaller secular frequencies $\nu_i$ in each
direction $i=x,y,z$. Due to the different time scales of secular
and micro motion, the latter can be neglected by time averaging
the equations of motion over a period of the r.f. drive. Under
these conditions, the effective time averaged potential assumes
the form:
\begin{equation}
\hat{\Phi} (\hat{x},\hat{y},\hat{z}) =  \frac{m }{2} \left(
\nu_x^2 \; \hat{x}^2 +  \nu_y^2 \; \hat{y}^2 +  \nu_z^2 \;
\hat{z}^2 \right)
\end{equation}
where  $\nu_x, \nu_y$ and $\nu_z$ are the secular frequencies in
the $x$, $y$ and $z$ axis of the trap and, for quadrupole traps as
the Paul trap, satisfy the relation $\nu_x + \nu_y = \nu_z$.

Thus the quantized c.m. motion of a single trapped ion in such an
oscillating electric quadrupole field can be effectively described
by a 3D harmonic oscillator with the Hamiltonian
\begin{equation}
\hat{H}_{cm}=\sum_{i=x,y,z} \hbar \nu_i \left( \hat{a}_i^{\dag}
\hat{a}_i + \frac{1}{2}\right)
\end{equation}
Here $\hat{a}_{x(y,z)}^{\dag}$ and  $\hat{a}_{x(y,z)} $ are the
creation and annihilation operators of vibrational quanta in the
mode corresponding to the motion along the $x(y,z)$  direction,
respectively. As usual the relation between
$\hat{a}_{x(y,z)}^{\dag}$ and  $\hat{a}_{x(y,z)} $ with the
position operator $\hat{x}$ and its conjugate momentum $\hat{p}_x$
is expressible by
\begin{equation}
\hat{x} = \sqrt{\frac{\hbar}{2m \nu_x}} \left(\hat{a}_x^{\dag} +
\hat{a}_x \right) \hspace{1cm} \hat{p}_x = i \sqrt{\frac{m \hbar
\nu_x }{2}} \left(\hat{a}_x^{\dag} - \hat{a}_x \right)
\end{equation}
Analogous relations hold  for the other two directions $y$ and
$z$.

\subsubsection{Interaction of single trapped ions with external
laser fields: the JC model}

%%\vspace{1cm}

We first consider the situation where an oscillating classical
electromagnetic field propagating along the $x$ direction
irradiates a single trapped ion which is constrained to move in
this direction in a harmonic well with frequency $\nu_x$. In more
detail we consider a uniform linearly polarized e.m. wave written
as
\begin{equation}
\hat{\bf E}(x,t)={\bf E}_0 e^{i (k \hat{x} - \omega_L t)} + h.c.,
\label{2.1}
\end{equation}
resonantly driving a single electronic transition between two
ionic internal states $\vert - \rangle$ and $\vert + \rangle$ . In
passing we note that the electric field is an operator acting on
the Hilbert space of the harmonic oscillator as a consequence of
the presence of the position operator $\hat{x}$ in the exponent.
The Hamiltonian describing the system under scrutiny is the
following
\begin{equation}
\hat{H}=\hat{H}_0+\hat{H}_{int} \label{2.2}
\end{equation}
with
\begin{equation} \hat{H}_0= \hbar \nu_x \left(
\hat{a}_x^{\dag} \hat{a}_x + \frac{1}{2}\right) + \hbar \omega_0
\hat{\sigma}_z  \label{2.3}
\end{equation}
and
\begin{equation}
\hat{H}_{int}= - {\bf \hat{d}} \cdot {\bf \hat{E} }(\hat{x},t)
\label{2.4}
\end{equation}
where ${\bf \hat{d}} = {\bf d} \hat{\sigma}_+ + h.c.$, ${\bf d}$
being the complex matrix element of the appropriate dipole
operator between the states $\vert + \rangle$ and $\vert -
\rangle$.

In the Rotating Wave Approximation (RWA) the interaction
Hamiltonian assumes the form
\begin{equation} \hat{H}_{int}= \hbar \Omega e^{i (k \hat{x} - \omega_L
t)} \hat{\sigma}_+ + h.a.  \label{2.5}
\end{equation}
where the coupling constant $\Omega$ is defined as
\begin{equation}
\Omega = - \frac{{\bf d} \cdot {\bf E }_0}{\hbar} \label{2.5a}
\end{equation}
Passing to the interaction picture with respect to $\hat{H}_0$ we
get
\begin{equation} \hat{H}^I_{int}= \hbar \Omega \left( e^{i \nu_x
\hat{a}_x^{\dag} \hat{a}_x t / \hbar} e^{i k \hat{x}} e^{-i \nu_x
\hat{a}_x^{\dag} \hat{a}_x t / \hbar}\right) e^{-i( \omega_L
-\omega_0)t} \hat{\sigma}_+ + h.a.  \label{2.6}
\end{equation}
It is useful to expand the exponential operator $e^{i k \hat{x}}$
in its normally ordered form:
\begin{equation}
e^{i k \hat{x}} \equiv e^{i \eta ( \hat{a}_x + \hat{a}_x^{\dag})}=
 e^{- \eta^2 /2} \sum_{m,l=0}^{\infty} \frac{(i
\eta)^{m+l}}{m!l!}{\hat{a}_x^{\dag} m} \hat{a}_x^l  \label{2.7}
\end{equation}
In equation (\ref{2.7}) we have introduced the Lamb-Dicke
parameter $\eta= \sqrt{\frac{\hbar k^2}{2 m \nu_x}}$. This
parameter is proportional to the ratio of the center of mass
uncertainty $\Delta x = \sqrt{\hbar/ 2m \nu_x} $ in its
vibrational ground state and the laser wavelength $\lambda=2 \pi /
k $. The physical situation in which
\begin{equation}
\eta \sqrt{\langle n_x \rangle +1} \ll 1, \label{s2.7a}
\end{equation}
where $\langle n_x \rangle$ is the mean number of oscillatory
quanta along the $x$ direction, is commonly referred to as the
Lamb-Dicke limit. In such a condition the ion couples with  the
laser behaving as a point-like object. Alternatively, $\eta$ can
be seen as the ratio between the photon momentum $\hbar k$ and the
ionic ground state momentum uncertainty $\Delta p_x = \sqrt{\hbar
m \nu_x /2} $. Thus, in the Lamb-Dicke regime, the effect on the
center of mass wavepacket due to momentum transfer during
absorption or emission of a photon is negligible. We wish to
underline that very often in literature the condition $\eta \ll 1$
is referred as Lamb-Dicke limit. This condition is less
restrictive than the one expressed by inequality (\ref{s2.7a}).
 In the following we will
also refer to $\eta \ll 1$ as the Lamb-Dicke condition but having
in mind that the mean number of oscillatory quanta have to be
small enough to satisfy (\ref{s2.7a}).

Inserting  equation (\ref{2.7}) into equation(\ref{2.6}) yields
\begin{equation}
\hat{H}_{int}^I(t)= \hbar \Omega \hat{\sigma}_+ e^{- \eta^2 /2}
\sum_{m,l=0}^{\infty} \frac{(i \eta)^{m+l}}{m!l!}{\hat{a}_x^{\dag}
m } \hat{a}_x^l e^{i[\omega_0 - \omega_L + (m-l) \nu_x]t} + h.c.
\label{2.8}
\end{equation}

This Hamiltonian consists of a sum of terms oscillating at
frequencies differing from $\omega_0 - \omega_L$ by multiples of
the vibrational frequency $\nu_x$. In the so called
resolved-sideband regime defined by the condition $\nu_x \gg
\Omega, \Gamma$, $\Gamma$ being the total linewidth of the
electronic transition, the different vibrational sidebands are
clearly resolved and it is possible to tune the laser in resonance
with a specific vibrational sideband. Moreover, in such a physical
situation, if the laser is tuned to the $k$th sideband, with $k >
0$ and $k < 0$ corresponding to red and blue detuning
respectively,
\begin{equation} \omega_L = \omega_0 - k \nu_x,
\end{equation}
we may neglect in equation (\ref{2.8}) all terms oscillating with
multiple vibrational frequencies.   As a consequence the effective
interaction hamiltonian in the interaction picture looses its time
dependence, becoming
\begin{eqnarray}
\hat{H}^I_{int} (t) \simeq \hat{H}_{int}^k &\equiv& \hbar \Omega
\hat{\sigma}_+ \hat{f}_k \left( \hat{a}_x^{\dag}\hat{a}_x, \eta
\right) \left( i \eta \hat{a}_x \right)^k + h.c.
 \hspace{1.5cm} k>0 \label{2.9} \\
\hat{H}^I_{int} (t) \simeq \hat{H}_{int}^k &\equiv& \hbar \Omega
\hat{\sigma}_+ \left( i \eta \hat{a}_x^{\dag} \right)^{-k}
\hat{f}_{-k} \left( \hat{a}_x^{\dag}\hat{a}_x, \eta   \right)  +
h.c. \hspace{1.5cm} k<0 \nonumber
\end{eqnarray}
The operator valued function $\hat{f}_k \left(
\hat{a}_x^{\dag}\hat{a}_x, \eta \right) $, with $k > 0$ and in its
normal ordered form, can be written down as follows \cite{matos1}
\begin{equation}
\hat{f}_k \left( \hat{a}_x^{\dag}\hat{a}_x, \eta   \right)= e^{-
\eta^2 /2} \sum_{l=0}^{\infty} \frac{(i
\eta)^{2l}}{(l+k)!l!}{\hat{a}_x^{\dag}l} \hat{a}_x^l
\hspace{1.5cm} k>0 \label{2.10}
\end{equation}
The interaction Hamiltonian given by equation (\ref{2.9}), due to
the presence of the operator $ \hat{f}_k \left(
\hat{a}_x^{\dag}\hat{a}_x, \eta   \right) $, represents a
nonlinear k-quantum Jaynes-Cummings model. There exist conditions
under which this Hamiltonian reduces to the well known JC
Hamiltonian and multiquantum JC Hamiltonian. To better understand
this point consider  the vibronic Rabi frequency defined as
\begin{equation}
\Omega_{n,n+k}=\Omega \langle n \vert \hat{f}_k \left(
\hat{a}_x^{\dag}\hat{a}_x, \eta   \right) \left( i \eta \hat{a}_x
\right)^k \vert n+k \rangle = (i \eta )^k \Omega e^{- \eta^2 /2}
\sqrt{\frac{n!}{(n+k)!}}L_n^{(k)}\left( \eta^2 \right).
\label{2.11}
\end{equation}
Equation (\ref{2.11}) shows a nonlinear dependence on $n$
departing from that characterizing the usual $k$-quantum JC model.
Moreover, due to the presence of the associated Laguerre
polinomials in equation (\ref{2.11}), there exist pairs of values
of $\eta$ and $n>0$ in correspondence of which the laser-atom
coupling practically vanishes. Let us consider the nonlinear
$k=1$-quantum JC interaction. In the Lamb-Dicke regime, the
operator $\hat{f}_1$ appearing in the Hamiltonian can be
simplified as $\hat{f}_1 \simeq 1$ and thus the vibronic Rabi
frequency assumes the form $\Omega_{n,n+1}\simeq i \eta \Omega
\sqrt{n+1}$  showing the typical behaviour of the standard JC
model. The same considerations hold for a generic value of $k$
leading us to the following conclusions. The light mediated
interaction between the quantized motion of the c.m. of an ion
moving in a Paul microtrap and its electronic degrees of freedom
shows a nonlinear character. This behaviour, traceable  back to
the specific adopted vibronic-coupling mechanism, is typical of
this system and does not have analogues in cavity QED quantum
models \cite{matos1,vogrew}. However, in the Lamb-Dicke regime,
that is when the c.m. wavepacket is well localized in comparison
with the laser wavelenght, the nonlinear multiquantum JC
interactions reduce to the multiquantum JC  models common in
quantum optics. Nonetheless, it is worth stressing that ionic
nonlinearities play, in general, an important role in the
experiments and very often need to be taken into account to
describe correctly the dynamics of the
system\cite{nist1,bibbiaNIST}.

We wish to underline that many experiments have been performed,
during the last few years, verifying the theoretical description
of the 1D Hamiltonian model given by equation(\ref{2.9})
\cite{nist1}. Moreover, a huge class of vibrational states such as
Fock states, coherent states, squeezed states \cite{nist1}  and
Schr\"{o}dinger cat like states \cite{nist3}, have been
experimentally realized and even the tomographic reconstruction of
the Wigner function of a vibrational Fock state has been
demonstrated \cite{nist4}. In most of the experiments the
electronic states of the ion are coupled by using two photon
stimulated Raman transitions through a third optical level as
indicated in Fig. \ref{fig:2} \cite{nist1,nist3,bibbiaNIST,nist4}.
For sufficiently large detunings $\Delta$ such that the third
intermediate state can be adiabatically eliminated, only two
electronic levels are involved in the dynamics. In this sense
stimulated Raman transitions are formally equivalent to a narrow
single photon transition \cite{nist1,bibbiaNIST}. Moreover they
recombine the advantages of strong optical electric field
gradients (allowing manipulation of the state of motion) and
microwave stability of the crucial difference frequency. It has
been shown that \cite{nist1,bibbiaNIST} the effective Hamiltonian
model in this case has the same form of the one given by equation
(\ref{2.9}), where now the coupling constant $\Omega $ is given by
$\Omega = \Omega_1 \Omega_2 / \Delta$, with $\Omega_{1(2)}=
\frac{{\bf d}_{1(2)} \cdot {\bf E }_0^{(1(2))}}{\hbar}$ and
$\Delta$ detuning from the third optical level, and $\eta =
\sqrt{\frac{\hbar k^2}{2 m \nu_x}}$ with ${\bf k = k_1 - k_2}$.
 We indicate with ${\bf d}_{1}$ and ${\bf d}_{2}$, ${\bf E }_0^{(1)}$ and ${\bf
E }_0^{(2)}$, ${\bf k_1}$ and ${\bf k_2}$  the dipole matrix
elements, laser amplitudes and   wave vectors of the two lasers
guiding the Raman transition respectively. As a consequence, the
Lamb-Dicke parameter, depends, in this case,  also on the relative
angle between the two incident lasers through the modulus of ${\bf
k}$. In this way it is possible at the same time to excite the ion
motion along one direction only and to manipulate easily the L-D
parameter simply by changing the relative directions of the beams
irradiating the ion.

Raman type configurations can be employed also to implement
vibrational transitions without affecting the electronic state of
the ion as shown in Fig.\ref{fig:3}. The form of the effective interaction
Hamiltonian in the interaction picture is, in this case, very
similar to the one given
 by equations (\ref{2.9}) and
(\ref{2.10}), but electronic inversion operators are absent:
\begin{eqnarray}
\hat{H}_{int}^k &=& \hbar \Omega  \hat{f}_k \left(
\hat{a}_x^{\dag}\hat{a}_x, \eta   \right) \left( i \eta \hat{a}_x
\right)^k + h.a. \hspace{1cm} (k>0) \label{2.12} \\
 \hat{H}_{int}^k
&=& \hbar \Omega \left( i \eta \hat{a}_x^{\dag} \right)^{-k}
\hat{f}_{-k} \left( \hat{a}_x^{\dag}\hat{a}_x, \eta \right)  +
h.a. \hspace{1cm} (k<0) \nonumber
\end{eqnarray}

We conclude this subsection  sketching briefly a very efficient
method, based on the so called quantum jumps technique, used in
the experiments for measuring the population of the electronic
states of the ion \cite{Qjumps}. It is worth noting that such a
quantity is the only one directly measurable for trapped ion
systems. The method essentially consists in coupling the ground
electronic level $\vert - \rangle$ to a third auxiliary level
$\vert r \rangle$ by means of a third laser beam, as shown in
Fig.\ref{fig:4}. The transition $\vert - \rangle \leftrightarrow
\vert r \rangle$ is chosen to be a dipole allowed one. In this
conditions, the presence of fluorescence detects the ion in the
electronic ground level and its absence in the excited level. In
other words, the detection of null fluorescence prepares the ion
in the excited state since after the measurement the quantum
system collapses in the eigenstate correspondent to the measured
outcome. With this method also if a modest number of scattered
photons are detected, the efficiency of the ability to
discriminate between the two states approaches $100 \%$.

\subsubsection{Generic bidimensional Hamiltonian models}

In this subsection we wish to present the extension of the 1D Hamiltonian model given by equation (\ref{2.9}) to
the case wherein two independent modes of motion of the ionic c.m., in the effective harmonic potential present
inside the trap, are excited. In the previous subsection we have assumed that the laser interacting with the ion
 propagates along the $x$-axis of the trap. When the laser wave vector lies on a plane of the trap, say
the $x$-$y$ plane, the dynamics couples the motion on that plane only. In this context, usually referred to as 2D
trapped ion, a derivation analogue to the one given in\cite{vog1} leads to the following expression for the general
effective Hamiltonian in the interaction picture
\begin{equation}
\hat{H}_{int}^{m_x,m_y}= \hbar \Omega e^{i \Phi}   \hat{g}_{m_x}
\left( \hat{a}_x^{\dag},\hat{a}_x, \eta_x \right)  \hat{g}_{m_y}
\left( \hat{a}_y^{\dag},\hat{a}_y, \eta_y \right)
\hat{\sigma}_+^{\epsilon} + h.c.  \label{2.20}
\end{equation}
with
\begin{eqnarray}
\hat{g}_{k} \left( \hat{a}_j^{\dag},\hat{a}_j, \eta_j \right)  &=& \left(i \eta_j \hat{a}_j^{\dag} \right)^{|k|} \hat{f}_{|k|} \left( \hat{a}_j^{\dag} \hat{a}_j, \eta_j \right) \hspace{1cm} (k \ge 0) \label{2.16} \\
\hat{g}_{k}  \left( \hat{a}_j^{\dag},\hat{a}_j, \eta_j \right) &=&
\hat{f}_{|k|} \left( \hat{a}_j^{\dag} \hat{a}_j, \eta_j \right)
\left(i \eta_j \hat{a}_j \right)^{|k|} \hspace{1cm} (k < 0)
\label{2.17}
\end{eqnarray}
where we have assumed the resonance condition $ \omega_L -
\omega_0 = m_x \nu_x + m_y \nu_y $ and we have used the property
$\nu_z=\nu_x+\nu_y$, valid for all quadrupole trapping potentials.
In equations (\ref{2.16}) and (\ref{2.17}) the operator function
$\hat{f}_{|k|}$ is defined by equation (\ref{2.10}). The parameter
$\epsilon$ is equal to 1 when internal transitions are involved
(Fig.\ref{fig:2}) and $\epsilon =0$ when the internal state is
unchanged (Fig. \ref{fig:3}). In the Lamb-Dicke regime, that is
assuming $\eta_x, \eta_y << 1$, one gets
\begin{eqnarray}
\hat{H}_{int}^{m_x<0,m_y<0}= \hbar \Omega'  \left( \hat{a}_x
\right)^{|m_x|}
\left( \hat{a}_y \right)^{|m_y|}  \hat{\sigma}_+^{\epsilon} + h.c.  \label{2.21a} \\
\hat{H}_{int}^{m_x>0,m_y<0} = \hbar \Omega' \left(
\hat{a}_x^{\dag} \right)^{|m_x|}
 \left( \hat{a}_y \right)^{|m_y|}  \hat{\sigma}_+^{\epsilon} + h.c.  \label{2.21b} \\
\hat{H}_{int}^{m_x<0,m_y>0} = \hbar \Omega'  \left( \hat{a}_x
\right)^{|m_x|}
 \left( \hat{a}^{\dag}_y \right)^{|m_y|}  \hat{\sigma}_+^{\epsilon} + h.c.  \label{2.21c} \\
\hat{H}_{int}^{m_x>0,m_y>0}= \hbar \Omega'  \left(
\hat{a}^{\dag}_x \right) ^{|m_x|}
\left(\hat{a}^{\dag}_y \right)^{|m_y|}  \hat{\sigma}_+^{\epsilon} + h.c.  \label{2.21d} \\
\end{eqnarray}
with $\Omega'= \Omega (i \eta_x)^{|m_x|} (i \eta_y)^{|m_y|}$.
Hamiltonian models similar to the ones expressed by equations
(\ref{2.21b}) and (\ref{2.21c}) are considered in
\cite{vog1,drobn1}. Both in \cite{drobn1} and in \cite{vog1} it is
shown that, in the case $\epsilon = 0$, an appropriate choice of
the detunings of the lasers leads to effective Hamiltonians
typical of the non linear optics context. We will analyze in more
detail some examples in Section II.

It is important to underline that the derivation of the effective
Hamiltonians  (\ref{2.20})  with the methods presented in
\cite{vog1,drobn1,bibbiaNIST,stein1}  is valid under the
assumption that the trap frequencies $\nu_x, \nu_y$ and $\nu_z$
are incommensurate. When this condition is violated, as for
example in the case of bidimensional isotropic traps for which
$\nu_x = \nu_y$, \lq \lq unwanted\rq \rq resonant terms may appear
in the effective interaction Hamiltonian in the interaction
picture. For isotropic traps, in order to avoid the appearance of
such terms, it is necessary to use more complex laser
configurations.

%It is important to underline that the case of isotropically
%confined ions is the one typically considered in proposals for
%generating bidimensional vibrational nonclassical states. Indeed,
%such proposals are all based on the time evolution of a certain
%ionic state under the action of a properly engineered effective
%Hamiltonian which, usually, is given in the interaction picture.
%It turns out that, in the case of isotropically confined ions, due
%to the presence of constants of motion, it is often easier to
%obtain the time evolution of the ionic state in the Schroedinger
%picture starting from the dynamics in the interaction picture.

\subsection{Dynamics}
\subsubsection{Trapped ions as quantum simulators} In this section
we describe the dynamics of some 2D trapped ion models studied in
literature over the last few years. As we have already pointed
out, initial studies on single trapped ions put into evidence the
formal analogy existing, under certain conditions, between these
systems and some nonlinear optical systems. As a consequence, when
these physical conditions are satisfied, the Hamiltonian models
given by equations (\ref{2.21a})-(\ref{2.21d}), can lead to
simulations of various devices of practical interest. As an
example, a (50/50) beamsplitter can be simulated. The action of
the beamsplitter on vibrational Fock states has been studied by
Gou and Knight in the case of 2D isotropically confined ions
\cite{MZinterfSU(2)}. The authors show that for these systems the
unitary operator describing the beamsplitter coincides with that
one describing rotations of the $x-y$ axes of the trap around the
$z$ axis in the real space. Similarly in \cite{nist2,bibbiaNIST} a
Mach-Zender interferometer acting on the two modes of oscillation
of a single trapped ion, has been considered. The analogy with the
Mach-Zender interferometer for bosons is that the two input modes
to the boson interferometer are replaced by the $x$ and $y$ modes
of ion oscillation.

Another example of the use of trapped ions to simulate the
dynamics of optical sistems is given in \cite{agarw1}. In this
paper, indeed, Agarwal and Banerji demonstrate that a single
bidimensionally confined ion provides a natural way of realizing
the quantum limit of second harmonic generation in a degenerate
optical parametric oscillator (OPO). They show that simulating the
quantum regime of OPO allows to demonstrate the existence of an
effect strictly quantum in nature.

\subsubsection{Motional nonlinearities} All the examples concerning
the use of trapped ions as quantum simulators rely on the
assumption that the Lamb-Dicke condition is satisfied. However,
when the spatial extension of the atomic wave function
representing the c.m. motion is no longer small compared with the
driving laser wavelenght, nonlinear effects having no counterpart
in standard nonlinear optics emerge. In \cite{vog1,vogrew} several
Raman-type excitations for inducing various kinds of nonlinear
interactions in the quantized motion of a trapped ion are
presented. It is there shown that standard effects can be
realized, including coherent displacement, Kerr nonlinearities and
parametric modes couplings. In the laser-assisted motional
dynamics, however, far from the Lamb-Dicke regime, additional
nonlinearities appear because of the effects on the atomic wave
function due to the driving laser waves. One of the consequences
of these nonlinearities consists in a nonlinear partitioning of
the phase space, the action of the time evolution being different
in neighboring phace-space zones \cite{vog1}.  To better
understand the origin of the phase space partioning let us
consider first, for semplicity, the single mode dynamics, where
only the motion in the $x$ direction is affected by the lasers
($\eta_y=\eta_z=0$). In this case the Hamiltonian is given by
equation (\ref{2.12}) with $ \hat{f}_k \left(
\hat{a}_x^{\dag}\hat{a}_x, \eta   \right)$ given by equation
(\ref{2.10}). If the Raman lasers are tuned to the first
vibrational sideband, $\omega_L = \omega_{L1} - \omega_{L2}=
\nu_x$, the structure of the time evolution operator shows some
formal resemblance to a nonlinearly modified coherent displacement
operator:
\begin{equation}
\hat{U}_{int}(t) = \hat{D} \left[ - \eta_x \Omega^{*}
\hat{f}_1(\hat{n}_x, \eta_x) t \right] =  exp \left[ - \eta_x
\Omega^{*} t \hat{a}_x^{\dag} \hat{f}_1(\hat{n}_x, \eta_x) + h.c.
\right] \label{s3.2}
\end{equation}
Note that for small values of the Lamb-Dicke parameter, $\eta_x
\ll 1$, the operator (\ref{s3.2}) may be replaced by the usual
displacement operator $\hat{D}(-\eta_x \Omega^{*} t/2)$.  Insight
into the nonlinear modifications of the displacement can be
obtained by considering the expectation value of the operator
function $\hat{f}_1(\hat{n}_x, \eta_x)$ in a coherent state $\vert
\alpha \rangle$: \begin{equation} \langle \alpha \vert
\hat{f}_1(\hat{n}_x, \eta_x) \vert \alpha \rangle = \left( \eta_x
|\alpha| \right)^{-1} J_1(2 \eta_x |\alpha|) e^{-\eta_x^2/2}
\label{s3.3} \end{equation} This function shows the oscillatory
behaviour typical of the Bessel function, where the zeros are of
special importance. Indeed these zeroes lead to the partitioning
of phase-space into different (radial) zones, with their
boundaries being defined by the zeros. In neighbouring phase-space
zones the actions of the time evolution appear to be significantly
different from each other. For example, one can see from numerical
calculations that in two adjacent zones a nonlinearly modified \lq
\lq displacement\rq \rq operator acts in opposite directions.
Consequently, a quantum state initially located on the boundary
between two zones will be split in two subspaces which eventually
give rise to quantum interferences.

When two or three dimensions are involved in the Raman induced
dynamics, the partioning effect appears in the phase space of each
motional degree of freedom. As a consequence, the coupling between
different motional modes will be strongly influenced by the
interplay of these nonlinear effects. So, as pointed out by
Wallentowitz and Vogel in \cite{vog1}, besides the feasibility of
realizing phenomena well known from linear optics in the motion of
bidimensionally trapped ions, these systems open up novel
possibilities for studying new nonlinear quantum phenomena.

\subsubsection{Quantum effects in the dynamics of bidimensionally
confined ions} In this subsection we discuss some quantum effects
recently brought to light in the dynamics of 2D isotropically
confined ions \cite{actamio,jmomio2}. Quantum phenomena similar to
the ones we are going to describe have been also studied in cavity
QED context. Nonetheless, in the context of trapped ions they
acquire a new physical meaning and they appear to be much easier
to verify experimentally.   Before describing the system under
scrutiny we want to underline that here, as well as in many of the
theoretical proposals we will describe in this review paper,  an
optical transition coupling directly the electronic states is
considered. In the experiments, however, two photon Raman
transitions are preferred. However, it is always possible to
substitute the direct coupling scheme with the two photon Raman
scheme obtaining the same effective Hamiltonian models
\cite{nist1}.

We consider a two-level ion  confined in a bidimensional isotropic
harmonic potential characterised by the trap frequency
$\nu_x=\nu_y=\nu$. It is well known that the bidimensional
harmonic oscillator may be associated to a generalised Schwinger
angular momentum  operator \mbox{$ {\bf \hat{J}} \equiv \left(
J_1, J_2, J_3 \right)$} as follows \begin{equation} \hat{J}_1 =
\frac{\hat{a}_x^{\dag} \hat{a}_y + \hat{a}_y^{\dag} \hat{a}_x}{2}
\hspace{1cm} \hat{J}_2 = \frac{ \hat{a}_x^{\dag} \hat{a}_y -
\hat{a}_y^{\dag} \hat{a}_x }{2i} \hspace{1cm} \hat{J}_3= \frac{
\hat{a}_x^{\dag} \hat{a}_x - \hat{a}_y^{\dag} \hat{a}_y}{2}
\hspace{1cm} \label{s3.4}
\end{equation}
It is worth stressing that these generalized angular momentum
operators acquire, in the context of 2D isotropically confined
ions, a direct and interesting physical meaning. The operator
$\hat{J}_3$ is indeed simply half the difference between the
number of vibrational quanta along the $x$ and $y$ axis of the
trap. The operator $\hat{J}_2$ is proportional to the
$z$-component of the ionic adimensional  angular momentum operator
$\hat{L}_z \equiv \frac{1}{\hbar} \left( \hat{x} \hat{p}_y -
\hat{y} \hat{p}_x \right) \equiv  -i \left( \hat{a}_{x}^{\dag}
\hat{a}_{y} - \hat{a}_{y}^{\dag} \hat{a}_x \right)=2 \hat{J}_2$.
Finally the operator $\hat{J}_1$ has the property of being
proportional, under certain conditions,  both to the quantum
covariance $C(\hat{x},\hat{y}) = \langle \hat{x}\hat{y} \rangle -
\langle \hat{x} \rangle \langle  \hat{y}  \rangle $  between the
position operators $\hat{x}$, $\hat{y}$ and to the quantum
covariance $C(\hat{p}_x,\hat{p}_y) = \langle  \hat{p}_x \hat{p}_y
\rangle - \langle  \hat{p}_x  \rangle \langle  \hat{p}_y  \rangle$
between the momentum operators $\hat{p}_x$, $\hat{p}_y$
\cite{actamio}.

It has been shown that by irradiating the ion with an appropriate
configuration of laser beams the physical system can be studied,
in the L-D limit and in the interaction picture, by the following
Hamiltonian model
\cite{pair,actamio,jmomio2,pramio,jmomio,rompmio}:
\begin{equation}
\hat{H}_{int}= \hbar \Omega' \left(
\hat{a}_x \hat{a}_y \hat{\sigma}_+ + h.a. \right) \label{s3.7}
\end{equation}
with $\Omega' = \Omega \eta^2 $. This Hamiltonian  is unitarily
equivalent to the last term of equation (\ref{hdsempl})
responsible for the parity effect presented in section II D 4, in
the context of cavity QED. In this case, however, Stark shift
terms present in equation (\ref{hdsempl}) are absent and the
parity dependent quantum effect discussed in section II D 4 is
much \lq \lq cleaner\rq \rq. Moreover it turns out that new
physical observable, having a direct physical meaning only in the
trapped ion context, show an interesting nonclassical behaviour.
In particular in \cite{actamio} it is shown that, starting from an
initial highly excited unidimensional Fock state $\vert
n_{\bar{x}} = N, n_{\bar{y}} = 0 \rangle$ with $N \gg 1$,  along
the direction $\bar{x}$ having an angle of $\pi / 4$ relative to
the $x$ axis, the time evolution of the mean value of the operator
$\hat{J}_1$ may be written as follows:
\begin{equation}
\langle  \hat{J}_{1 }(t) \rangle  \simeq
\frac{N}{2} \cos^{N-1} \left( \frac{\Omega' t}{N} \right)
\label{s3.8}
\end{equation}
From equation \ (\ref{s3.8}) it is easy to deduce that at the time
instant $t_N=  \frac{\pi N}{\Omega'}$ one gets $ \langle
\hat{J}_{1 }(t) \rangle  \simeq (-1)^{N-1} \frac{N}{2} $. This
result clearly evidences that the system under scrutiny possesses
an inherent peculiar sensibility to the parity of the initial
total number $N$ of vibrational quanta. There exists indeed a
$N$-dependent instant of time $t_N$ at which $\langle  \hat{J}_{1
}(t) \rangle$ assumes values crucially related to the evenness or
oddness of $N$. In more detail, $ \langle  \hat{J}_1 (t_N) \rangle
$ is equal to $-N/2$ when $N$ is even and $+N/2$ when $N$ is odd.
Fig.\ref{fig:5} shows the time evolution of the quantity $\langle
\hat{J}_{1 }(t) \rangle$. Remembering the link between $\langle
\hat{J}_1 (t) \rangle$ and the quantum covariances $C(\hat{X},
\hat{Y})$ and $C(\hat{P}_{x}, \hat{P}_{y})$ and taking into
consideration that the maximum (minimum) mean value of $\hat{J}_1$
is $N/2$ ($-N/2$), this result can be readily interpreted in the
following way. The dynamics of the system under scrutiny is
characterised by the existence of a N-dependent instant of time at
which the ionic oscillatory motions along the  $x$ and $y$ axis
are maximally or minimally correlated in position and momentum in
dependence on the oddness or evenness of $N$ respectively.

Another interesting quantity displaying a parity dependent
temporal evolution is the variance of the operator $\hat{J}_2$
which is proportional to the variance of the axial orbital angular
momentum operator $\hat{L}_z$. In a system identical to the one
previously described, the mean value of $\hat{J}_2$ is null at any
time instant $t$, while at the time instant $t_N/2 = \pi N /2
\Omega$ the square of the variance assumes the value
\cite{jmomio2}
\begin{equation}
(\Delta J_2)^2(t_N/2) \simeq \left(1-(-1)^N \right) \frac{N^2}{8}
+O(N) \label{s3.11}
\end{equation}
This means that $ (\Delta J_2)^2(t_N/2) $ is of the order of $N\gg
1$ when the initial number of vibrational quanta is even  and of
the order of $N^2$ when it's odd. In other words the variation of
one vibrational quantum only in the initial conditions of the
system, quite drastically modifies the dispersion around zero of
the measured values of the ionic angular momentum $z$-component at
a subsequent specific instant of time. In \cite{pramio} we have
shown that at this time instant, $t_N/2$, the vibrational and
electronic degrees of freedom of the ion are disentangled for odd
$N$ and maximally entangled for even $N$. At $t=t_N/2$, indeed,
the probability of finding the ion in its ground electronic state
is $1/2$ or $1$ in correspondence to $N$ even or odd respectively.
We have also checked the robustness of this parity dependent
entanglement effect against the non-dissipative decoherence due to
laser intensity fluctuations and imperfect initial state
preparation finding that, for values of the laser parameters
currently used in the experiments, the nonclassical effect,
althought attenuated, is still present \cite{jobmio,caprimio}.

\subsection{Generation of nonclassical states}
\subsubsection{Synthesis of arbitrary states of two-dimensional
vibrational motion} As we have already mentioned in section III A,
trapped ions are exceptionally well suited systems to study
fundamental aspects of quantum theory. By irradiating the ion with
laser fields coupling its internal and external degrees of
freedom, a single ion can be cooled to the vibrational ground
state of the trap and its motion can be manipulated coherently to
generate nonclassical states of motion. The ability to synthesize
arbitrary motional states and to control their evolution not only
allow to verify experimentally  fundamental concepts of quantum
mechanics, but also opens new horizons for potential technical
applications such as quantum information processing. For these
reasons, during the last few years, several proposals for
engineering arbitrary nonclassical states of the 1D, 2D or 3D
oscillatory motion of the ionic center of mass have been presented
(see \cite{vogrew} and references therein for a review). In what
follows we concentrate our attention on bidimensional engineering
schemes.

The first scheme for preparing an arbitrary motional state of a
bidimensionally trapped ion was proposed by Gardiner, Cirac and
Zoller \cite{gard1} as a key prerequisite for quantum measurement
of an arbitrary motional observable. The method, conceptually
similar to the one proposed by Law and Eberly in the context of
cavity QED \cite{laweb}, consists in preparing a two-mode target
state of the form
\begin{equation} \vert \Psi_{target} \rangle =
\sum_{n_x=0}^{N_{max}} \sum_{n_y=0}^{M_{max}} c_{n_x,n_y} \vert
n_x, n_y \rangle, \label{s4.1}
\end{equation}
starting from the initial vibrational and electronic ground state
$\vert n_x=0, n_y=0, - \rangle$. Such a state is generated by
means of an appropriate sequence of laser pulses used to
coherently distribute the initial amplitude to form a desired
arbitrary superposition.

Although conceptually elegant, the method proposed in \cite{gard1} is difficult to implement since the number of
laser pulses necessary for preparing the bimodal target state depends exponentially  on the dimensionality of the
subspace of the Fock space in which the target state is embedded. Ideal synthesis of motional quantum states
assumes indeed negligible dissipation effects and perfect control of areas and phases of the laser pulses and
switching times. Fluctuations of these quantities cause some technical noise affecting the efficiency of the
schemes. For these reasons it is desirable to obtain preparation schemes based on a number of laser pulses as small
as possible. In this spirit, during the last few years, novel approaches allowing to reduce consistently the number
of operations required for synthesizing desired target states have been proposed \cite{kneer1,drobn2,zhen1}. In
particular very recently Zheng presented a method for preparing arbitrary bidimensional vibrational states of a
trapped ion of the form (\ref{s4.1}) requiring only $(M_{max}+2)(N_{max}+1)$ laser pulses \cite{zhen1}. It is worth
noting, however, that a state of the form (\ref{s4.1}) involves $(M_{max}+1)(N_{max}+1)$ complex coefficients, each
controlled by the amplitude and phase of a laser pulse. Thus it should exist, in principle, a better approach
requiring only $(M_{max}+1)(N_{max}+1)$ operations \cite{zhen1}.

All the procedures we have described till now concern the
generation of arbitrary bimodal oscillatory states of the c.m.
motion of an ion confined in a trapping potential. They rely on
the possibility of engineering suitable interaction Hamiltonians,
generally involving the two mode bosonic operators and the atomic
ones, to control the evolution from the initial state to the
desired target state.  It has been shown that, like in the case of
bimodal cavity fields, such bimodal interactions can also be used
to synthesize more efficiently arbitrary  unidimensional
vibrational states. In a recent proposal, Hladky, Drobny and Buzek
show how to engineer an arbitrary unitary operator $\hat{V}$,
transforming any state $\vert \Psi \rangle $ of a single bosonic
mode, e.g. a 1D vibrational state of a trapped ion along the $x$
direction, to another state $\vert \Psi' \rangle = \hat{V} \vert
\Psi \rangle$ \cite{hladk1}. The realization of the unitary
transformation $\hat{V}$ is based on an enlargement of the Hilbert
subspace of the system. In the case under scrutiny the ionic c.m.
mass mode of oscillation along $x$ becomes entangled with
auxiliary degrees of freedom (ancilla) which are represented by
the vibrational mode along the $y$ direction and three internal
electronic levels of the ion. The physical realization of the
desired operator $\hat{V}$ consists again of a sequential
switching of laser fields irradiating the ion, the correspondent
interaction Hamiltonians being typical and quite easy to realize
in the trapped ion context. The overall algorithm involves a
conditional measurement of an internal ionic state, thus the
process is not unitary. It is, however, universal since the
probability of the conditional measurement does not depend on the
initial vibrational state.

\subsubsection{Bimodal generalized coherent states}
In this subsection we will review some recent proposals for realizing, in the context of trapped ions,  generalized
coherent states (GCS)\cite{GCS} such as pair coherent states, SU(1,1) coherent states and SU(2) coherent states.
Over the last several years there has been much discussion on GCS  of the quantized electromagnetic field since
they exhibit strong nonclassical properties such as sub-Poissonian statistics, squeezing, violation of
Cauchy-Schwarz inequality and so on \cite{SU11}. Unfortunately they are not very easy to realize in practice in the
optical context. We have already seen, however, that it is possible to imagine rather simple mechanisms by which a
great variety of effective Hamiltonians can be engineered in the trapped ion context. Such interaction models can
be used, as we shall see in the following, to generate several classes of vibrational states including generalized
coherent states. There exist basically two  main ways to generate a given vibrational state with trapped ions. The
first way consists in the coherent evolution due to an appropriately engineered interaction Hamiltonian, starting
from a given initial state. Eventually a conditional measurement can be performed to project the ionic vibrational
state in the desired target state. The second way takes into account also the effects of decoherence due to
spontaneous emission in the process of state generation. An advantage of this last method is that the desired
states are obtained as steady state solutions of the master equation containing dissipative terms due to
spontaneous emission. These target states are called dark states because their generation experimentally
corresponds to the extinction of fluorescence by the ion. Whenever a given class of states is defined by an
eigenvalue equation, it is possible to imagine, in the trapped ion context, a configuration able to realize the
dark target state starting from a proper initial condition of the system. To better understand this point we remind
that, as the damping of vibrational quanta can be significantly suppressed in an ion trap, the dominant decay
process is the spontaneous emission from the two-level ion. Thus the time evolution of the system, in the
interaction picture, can be described by a density operator $\hat{\rho} $ obeying the master equation
\begin{equation}
\frac{d \hat{\rho}}{dt} = - \frac{i}{\hbar} \left[
\hat{H}_{int}^{I} , \hat{\rho} \right]+ \frac{\Gamma}{2} \left( 2
\hat{\sigma}_- \hat{\varrho} \hat{\sigma}_+ - \hat{\sigma}_+
\hat{\sigma}_- \hat{\rho}- \hat{\rho} \hat{\sigma}_+
\hat{\sigma}_- \right) \label{s4.6}
\end{equation}
where $\Gamma$ is the spontaneous decay rate of the excited ionic
state and $\hat{\varrho}$ is the modified density operator
\begin{equation} \hat{\varrho}= \frac{1}{4}\int_{-1}^{1}
\int_{-1}^1 du \; dv \, W(u,v) e^{ik(u \hat{x} + v\hat{y})}
\hat{\rho} e^{-ik(u \hat{x} + v\hat{y})} \label{s4.7}
\end{equation}
accounting for the momentum transfer in the $x-y$ plane due to
spontaneous emission. In equation (\ref{s4.7}) $W(u,v)$ describes
the angular distribution of the spontaneous emission. The steady
state solution of equation (\ref{s4.6}) has the form $\hat{\rho}_S
= \vert - \rangle \vert \Psi \rangle \langle \Psi \vert \langle -
\vert$, where $\hat{\rho}_S $ satisfies the restriction $\left[
\hat{H}_{int}^{I} , \hat{\rho}_S \right]=0 $. If the interaction
Hamiltonian has the form
\begin{equation}
 \hat{H}_{int}^I=  \hbar \Omega \left[
\hat{A}_{vib} - \epsilon \right] \sigma_+ + h.a. \label{dipiu}
\end{equation}
, where $\hat{A}_{vib}$ is a generic  vibrational operator, then the stationary state of the system will satisfy
the eigenvalue equation $\hat{H}_{int}^I \vert \Psi \rangle = \epsilon \vert \Psi \rangle$.

\vspace{1cm}

{\it a) SU(2) coherent states}

We have seen in subsection III B 1 that a Mach-Zender
interferometer can be simulated  with 2D trapped ion systems. An
SU(2) interferometer is a peculiar kind of Mach-Zender
interferometer of particular relevance in quantum optical theory.
The vibrational SU(2) states are anticorrelated bimodal states
formally identical to the photon states on the output ports of a
SU(2) interferometer with number states inputs. As a consequence a
way to generate vibrational SU(2) states is to prepare an initial
bimodal Fock state and then to realize an effective Hamiltonian
\lq \lq simulating\rq \rq the action of a SU(2) interferometer. In
\cite{MZinterfSU(2)} it is shown that the mode-mixing process on
the output ports of the interferometer is replaced, in the context
of trapped ions, by a real-space rotation in the $x$-$y$ plane.
This means, in other words, that a bimodal Fock state excited
along the $x-y$ axis is an SU(2) state in the coordinate system
rotated of an angle $\theta$ with respect to $x-y$. We wish to
remind that, as seen in subsection III B 3, an isotropic
bidimensional harmonic oscillator may be associated to a
generalized Schwinger angular momentum operator $\hat{J}$ whose
components are defined in equation (\ref{s3.4}). The two mode Fock
state $\vert n_x, n_y \rangle$ can be described as a pseudo
angular momentum state $\vert j, m \rangle$ which is the common
eigenstate of the angular momentum operators $\hat{J}^2$ and
$\hat{J}_3$ with correspondent eigenvalues
$j=\frac{1}{2}(n_x+n_y)$ and $m =\frac{1}{2}(n_x-n_y)$. The
generalized SU(2) coherent states are defined by the action of the
generalized SU(2) displacement operator as follows
\begin{equation}
\vert \tau, j \rangle = e^{\left( \beta \hat{J}_+ - \beta^*
\hat{J}_- \right)} \vert j, -j \rangle \label{s4.3}
\end{equation}
where $\hat{J}_+ = \hat{J}_1 + i \hat{J}_2 $, $\hat{J}_- =
\hat{J}_1 - i \hat{J}_2 $, $\beta = \alpha e^{(-i \gamma)}$ ($0
\le \alpha \le \pi, \; 0 \le \gamma \le 2 \pi$) and $\tau =
\tan(\alpha/2) e^{(-i \gamma)}$. Reminding that the operator
$\hat{J}_2$ is proportional to the $z$ component of the orbital
angular momentum operator generating rotations in the $x-y$ plane,
it is not difficult to convince oneself that the method proposed
in \cite{MZinterfSU(2)} for generating SU(2) states allows also to
engineer \lq \lq real\rq \rq SU(2) coherent states, that is states
$\vert \tau, j \rangle$ with $\tau$ real.

Very recently another method for generating dark SU(2) states of the bidimensional oscillatory motion of a single
trapped ion has been proposed \cite{zsolt}. Such states are  eigenstates of both the total number of vibrational
quanta and the vibrational operator $\hat{A}_{vib}=\hat{a}_x \hat{a}^{\dag}_y + h.a.$. Due to the degeneration of
their eigenvalue spectrum, however, the dark states obtained as steady state solutions of the master equation
(\ref{s4.6}) would be incoherent superpositions of SU(2) states. It is shown in \cite{zsolt} that, for large values
of the L-D parameter, motional nonlinearities come into play removing the eigenvalues degeneration. In these
circumstances it is possible to show numerically that the steady state solutions approximate well SU(2) states
\cite{zsolt}.

\vspace{1cm}

{\it b) Dark pair and SU(1,1) coherent states}

Let us now discuss two similar methods for generating dark pair
coherent states and SU(1,1) coherent states \cite{pair,SU11}. We
firstly review the procedure for engineering pair coherent states
\cite{pair}.

The pair coherent states $\vert \xi; q \rangle$ are defined as the
eigenstates of both the pair annihilation operator $\hat{a}_x
\hat{a}_y$ and the number difference $\hat{Q}=\hat{n}_x-\hat{n}_y$
and, in the Fock state basis have the form:
\begin{equation} \vert \xi; q
\rangle = N_q \sum_{l=0}^{\infty} \frac{\xi^l}{\sqrt{l! \;
(l+q)!}} \vert l+q,l \rangle \label{s4.4}
\end{equation} where $N_q=\left[ |\xi|^{-q}I_q(2|\xi|) \right]$ is the normalization constant ($I_q$ is the
modified Bessel function of the first kind of order $q$). It is
then clear that, by engineering an Hamiltonian model of the form
\begin{equation}
\hat{H}_{int}^{I}= \hbar \Omega  \left[ \hat{a}_x \hat{a}_y - \xi
\right] \hat{\sigma}_+ + h.a. \label{s4.5}
\end{equation}
the dark state obtained is an eigenstate of $\hat{a}_x \hat{a}_y$
with eigenvalue $\xi$. This means that, since the number
difference operator $\hat{Q}$ is a constant of motion in processes
involving simultaneous pair annihilation or creation, starting
from an eigenstate of $\hat{Q}$, like $\vert \Psi_0 \rangle =
\vert q,0 \rangle$, one obtains as dark steady state the pair
coherent state $\vert \xi, q \rangle$.

A similar technique is used in \cite{SU11} to create one mode and
two mode SU(1,1) intelligent states of the vibrational ionic
motion. SU(1,1) intelligent states are a class of SU(1,1) coherent
states defined by the property that an SU(1,1) uncertainty
relation is equalized \cite{cavityqed2}. It can be shown
\cite{merz} that these states are solutions of the following
eigenvalue problem
\begin{equation} \left( \alpha \hat{a}_x \hat{a}_y + \beta
\hat{a}_x^{\dag} \hat{a}_y^{\dag} \right)\vert \Phi \rangle =
\zeta \vert \Phi \rangle \label{s4.8}
\end{equation}
Moreover, they are eigenstates of the number difference operator
$\hat{Q}$. Thus, preparing the system in an eigenstate of
$\hat{Q}$ and irradiating the ion with a configuration of lasers
realizing the Hamiltonian model
\begin{equation}
\hat{H}_{int}^{I}= \hbar \left[ \Omega \hat{a}_x \hat{a}_y +
\bar{\Omega}  \hat{a}_x^{\dag} \hat{a}_y^{\dag} - \zeta \right]
\hat{\sigma}_+ + h.a. \label{s4.9}
\end{equation}
the steady state solution, indicated by extintion of fluorescence
emission by the ion, is an intelligent SU(1,1) state.

\vspace{1cm}

{\it c) Generation of pair coherent states via QND measurements}

To conclude this subsection we wish  to describe a recent
alternative technique for generating pair coherent states
\cite{lasphmio,chinmio}. The scheme is based on a Quantum Non
Demolition (QND) measurement procedure and allows actually to
engineer several classes of vibrational states \cite{prawang}. The
ion is irradiated by two laser beams propagating along the $x$ and
$y$ axis of the trap and realizing, in the L-D limit and in the
interaction picture an interaction of the form
\cite{lasphmio,chinmio,prawang,gerr2}
\begin{equation}
\hat{H}_{int}^{I}= |\Omega _{Lx}-\Omega _{Ly}-\chi
(\hat{n}_x-\hat{n}_y)| \hat{\sigma}_+ + h.a.=|\Omega _{Lx}-\Omega
_{Ly}-\chi \hat{Q}| \hat{\sigma}_+ + h.a. \label{s4.11}
\end{equation} where
$\Omega_{Lx}= d E_x / \hbar$, $\Omega_{Ly}=d E_y / \hbar$ and we
have assumed the condition  $\Omega _{Lx}\eta _x^2=\Omega
_{Ly}\eta _y^2\equiv \chi ,$ with $\Omega _{Lx}\neq \Omega _{Ly}.$
The nonlinear JCM of equation (\ref{s4.11}) is exactly solvable
and it is not difficult to prove that, if the system is initially
prepared in a state of the form $\vert n_x, n_y, - \rangle$, then
Rabi oscillations of the electronic occupation occur, the
vibrational state remaining unchanged:
\begin{equation} \vert \Psi _-(t)\rangle =\cos \left[ \Omega (n_x,n_y)t
\right]\vert n_x,n_y,- \rangle- i \sin \left[\Omega (n_x,n_y )t
\right] \vert n_x,n_y,+ \rangle,  \label{s4.12}
\end{equation}
with $\Omega (n_x,n_y) = |\Omega_{Lx}-\Omega _{Ly}-\chi q|$ and
$q=n_x-n_y$. We have seen in section III A that, in a typical
trapped ion experiment, a measurement of the internal electronic
state is realized coupling the ground state with a third level by
an auxiliary laser beam \cite{Qjumps}. If a null fluorescence is
detected then we may claim with certainty that the ion is not in
its ground state. In other words, this detection process prepares
the ion in the excited state. Suppose now that at $t=0$ the ion is
prepared in its electronic ground state whereas the vibrational
motion is in a generic superposition of two-mode Fock states.
Equation (\ref{s4.12}) suggests a simple protocol to generate a
prefixed bimodal Fock state $\vert m_x,m_y \rangle $ provided that
the probability of finding at $t=0$ the center of mass motion in
the state $\vert m_x,m_y \rangle $ is different from zero. In
order to generate this state, we imagine to measure several times
the internal electronic state at time instants $t_i$ appropriately
chosen.

The first measurement is performed at the time $t_1$ which
satisfies the condition $\Omega (m_x,m_y )t_1=(2l_1+1)\pi /2,$
$(l_1=0,1,2,\cdots )$. If the null fluorescence is detected,
according to equation (\ref{s4.12}), the vibrational state $\vert
m_{x,}m_y \rangle$ keeps the original probability unchanged
whereas the other components present at $t=0$ decrease their
probabilities. In other words, after the renormalization of the
state, the probability of the state $\vert m_{x,}m_y \rangle $ is
increased with respect to the other two-mode Fock states. This
process can be obviously iterated  measuring the null fluorescence
at time instants $t_k$ $(k=2,3,\cdots )$ defined by the condition
$\Omega (m_x,m_y )t_k=l_k\pi ,$ $(l_k=1,2,3,\cdots )$.
It is easy to convince oneself that, following this procedure, the prefixed state$%
\vert m_{x,}m_y \rangle$ can be generated. Remembering the Fock
state expansion of the pair coherent states, given by equation
(\ref{s4.4}), and noting that all the states $\vert n+q, n
\rangle$ share, in the system under scrutiny, the same Rabi
frequency $\Omega (m_x=n+q,m_y=n )$, it is possible to persuade
oneself that the procedure described before may be used in our
context to engineer pair coherent states. It turns out that,
starting from a bimodal coherent state $\vert \alpha, \beta
\rangle$ it is possible to generate the pair coherent state $\vert
\xi = \alpha \beta, q \rangle$ with fairly good efficiency. As an
example, Fig.\ref{fig:7} shows the probability distribution
relative to the vibrational state $\vert \alpha \beta, 2 \rangle$
in correspondence to $|\alpha |^2=2.5$ and $|\beta |^2=1.5$, and
$\,\chi/(\Omega _{Lx}-\Omega _{Ly})=0.0049$. It is worth stressing
that this quantum state manipulation procedure requires several
successful null fluorescence detections in order to generate the
desired state. In the case of the example of Fig.\ref{fig:7} the
probability of success is $P=0.172$, which is already a
significant value from an experimental point of view.

\subsubsection{Schr\"{o}dinger cat-like states} Since the famous
cat thought experiment suggested by Schr\"{o}dinger in 1935
\cite{schroedinger}, superpositions of macroscopically
distinguishable quantum states, also known as Schr\"{o}dinger cat
states, have become a longstanding example of the peculiarities
which occur in the interpretation of quantum reality. To explore
and gain insight into the fundamental issues of quantum theory, a
number of schemes have been proposed for the realization of such
states. However, the generation of cat states and the detection of
their quantum properties is an extremely hard tool since
decoherence effects due to the interaction with the external
environment tend to transform these quantum superpositions into
statistical mixtures destroying their coherence properties. For
this reason single trapped ions have shown to be very well suited
to the generation of Schr\"{o}dinger cat states, since, as we have
already pointed out, dissipative effects which are inevitable from
cavity damping in the optical or microwave regime, can be
significantly suppressed for the ion motion due to the extremely
weak coupling between the vibrational modes and the external
environment. The archetype of a Schr\"{o}dinger cat states is
given by the superposition
\begin{equation}
 \vert
\Psi \rangle_{cat} = N \left( \vert \alpha \rangle + e^{i \phi }
\vert - \alpha \rangle \right), \label{s4.12a}
\end{equation}
where $\alpha$ is a coherent state of a single mode quantized
field and $N$ is a normalization constant. These states are
referred to as even, odd and Yurke-Stoler coherent states when
$\phi=0, \pi$ and $\pi/2$ respectively \cite{yurke-stoler}.
Schr\"{o}dinger cat states of this type have been recently
realized for a trapped $^9Be^+ $ ion \cite{nist3}. With an average
of about nine vibrational quanta, wave packets of maximum
separation of about 83 nm, significantly larger than the size of a
single component wave packet, about 7 nm, were obtained.

\vspace{1cm}

{\it a) Bimodal Schr\"{o}dinger cat states}

In this subsection we review some proposals for generating
Schr\"{o}dinger cat-like states of the bidimensional oscillatory
motion of a single trapped ion. The first generation scheme we
consider \cite{gerr2} allows to engineer cat-like states which are
the generalization to two modes of motion of the states
experimentally realized in \cite{nist3}. The system considered is
an isotropically confined ion irradiated by laser beams in a
configuration similar to the one described in the previous
paragraph and realizing the effective Hamiltonian (\ref{s4.11}).
The only difference is that here the Rabi frequencies of the two
lasers are assumed to be equal $\Omega_{Lx}= \Omega_{Ly}=\Omega$.
With this restriction, the interaction Hamiltonian in the
interaction picture assumes the form
\begin{equation}
\hat{H}_{int}^I= - \chi  (\hat{n}_x-\hat{n}_y) \left(
\hat{\sigma}_+ + \hat{\sigma_-} \right) \label{s4.13}
\end{equation}
As shown in \cite{gerr2}, if the initial state is a bimodal
coherent states $\vert \alpha \rangle \vert \beta \rangle \equiv
\vert \alpha, \beta \rangle$ and the ion is in its ground internal
state, after a time $t > 0$ the state of the system evolves to
\begin{equation}
\vert \Psi (t) \rangle = \vert - \rangle \vert S_+ \rangle + \vert
+ \rangle \vert S_- \rangle \label{s4.14}
\end{equation}
where
\begin{equation}
\vert S_{\pm} \rangle = \frac{1}{2} \left( \vert \alpha e^{i
\phi/2}, \beta e^{-i \phi/2} \rangle \pm \vert \alpha e^{-i
\phi/2}, \beta e^{i \phi/2} \rangle \right)
\end{equation}
with $\phi = 2 \chi t$. The states $\vert S_{\pm} \rangle$ are
examples of two-modes cat states of the even and odd type and may
be characterized by oscillatory distribution of the vibrational
quanta and by strong nonclassical correlations between the modes.
It is important to underline the possibility of detecting the
entanglement between the modes, which is reflected in the
interference between the two components of $\vert S_+ \rangle$.
This can be experimentally realized in the same way as in the one
mode case, that is by measuring the probability $P_-(\phi)$ of
finding the ion in its internal ground state. Indeed $P_-(\phi)$
is given as (with $\alpha$ and $\beta $ real for simplicity)
\begin{equation}
P_-(\phi)= \langle S_+ \vert S_+ \rangle = \frac{1}{2} \left\{
1+e^{-(\alpha^2+ \beta^2)(1-\cos(\phi))} \cos \left[\left(
\alpha^2 - \beta^2 \right) \sin \phi \right] \right\}
\label{s4.15}
\end{equation}
and is sensitive both to the sum and to the difference of the
squares of the amplitudes of the coherent states.

\vspace{1cm}

{\it b) Mesoscopic superpositions of pair coherent states}

Till now we have been dealing with quite direct generalization of
the one mode cat state (\ref{s4.12a}) to two dimensions. There has
been however much interest also in the superpositions of
macroscopically distinguishable bimodal states such as pair
coherent states, SU(2) coherent states, squeezed states and so on.
These superpositions are usually referred to as Schr\"{o}dinger
cat-like states. Gou, Steinbach and Knight \cite{goupair}  present
a scheme for generating mesoscopic superpositions of two pair
coherent states separated in phase by $\phi$:
\begin{equation}
\vert \xi; q, \phi \rangle = N_{\phi} \left( \vert \xi; q \rangle
+ e^{i \phi} \vert - \xi; q \rangle \right) \label{s4.16}
\end{equation}
In equation (\ref{s4.16}) $N_{\phi}$ is a normalization constant
and the pair coherent states $\vert \xi; q \rangle$ are defined by
equation (\ref{s4.4}).  It is easy to verify that the states
$\vert \xi; q, \phi \rangle$ are eigenstates of the operator
$(\hat{a}_x \hat{a}_y)^2$ with eigenvalue $\xi ^2$. The scheme for
generating such states is a generalization of the one proposed by
the same authors for engineering pair coherent states and
discussed in the previous paragraph. Also in this case the desired
target state is obtained as steady state solution of a master
equation describing the effect of spontaneous emissions on a
system represented in terms of the following effective interaction
Hamiltonian in the interaction picture:
\begin{equation}
\hat{H}_{int}^I = \hbar \Omega \left[ \left( \hat{a}_x \hat{a}_y
\right)^2 - \xi^2 \right] \hat{\sigma}_- + h.a. \label{s4.15a}
\end{equation}
For general initial states of motion the oscillatory steady state,
detected by the extinction of the fluorescence emitted by the ion,
is a statistical mixture. However, the authors show that, if the
initial state of motion of the ion is prepared in the Fock state
$\vert q,0\rangle$ ($\vert q+1,1\rangle$), then the vibrational
steady state is described by an even (odd) pair coherent state.

\vspace{1cm}

{\it c) Mesoscopic superpositions of SU(2) coherent states}

Let us now consider some recent proposals for generating Schr\"{o}dinger cat like states based on conditional
measurements of the internal state of the ion. The first scheme we wish to describe is aimed at creating
superpositions of two $SU(2)$ coherent states separated in phase by $\pi$ \cite{pramio,aipmio}. The system
considered has been already described in subsection III B 3. We have there seen that a single bidimensionally and
isotropically confined ion, driven by an appropriate configuration of laser beams (see equation(\ref{s3.7})) and
prepared in an initial vibrational Fock state along the direction $\bar{x}$ with an angle $\pi /4 $ relative to the
$x$ axis, exhibits a temporal behaviour strongly sensitive to the parity of the initial total number of vibrational
quanta.  It is shown in \cite{pramio} that, if the ion is initially prepared in the vibrational state $\vert
n_{\bar{x}} = N, n_{\bar{y}} = 0 \rangle$ and in its electronic ground state, then the system dynamics, governed by
the Hamiltonian (\ref{s3.7}), is characterized by the existence of instants of time at which the conditional
measurement of the electronic ground state projects the c.m. motion into a superposition of two macroscopically
distinguishable SU(2) coherent states. In more detail, if the initial total number of vibrational quanta $N \gg 1$
is even, then at $t= t_e/2 = \pi N / 4 \Omega '$ the following superposition is generated
\begin{equation}
\vert \psi \left( t_e/2 \right) \rangle = \frac{1}{\sqrt{2}}
\left( \vert \tau = 1, j = N/2 \rangle + (-1)^{ \frac{N}{2} }
\vert \tau = - 1, j = N/2 \rangle \right) \label{s4.17}
\end{equation}
whereas, if $N$ is odd,  the state generated at $t=t_o \simeq
t_e/2 $ is
\begin{equation}
\vert \psi (t_o) \rangle = \frac{1}{\sqrt{2}} \left( \vert \tau =
i, j = N/2 \rangle - i (-1)^{\frac{N+1}{2} } \vert \tau = - i, j =
N/2 \rangle \right) \label{s4.18}
\end{equation}
Equations (\ref{s4.17}) and (\ref{s4.18}) show that the properties
of the state generated following the procedure envisaged in
\cite{pramio}, strongly depend on the parity of the total initial
number $N$ of vibrational quanta. In other words this means that,
for even $N$,  the state $\vert \psi \left( \frac{t_e}{2} \right)
\rangle $, defined by equation\ (\ref{s4.17}), has the form of an
even (odd) SU(2) coherent state \cite{gerrgro} , if $\frac{N}{2}$
is even (odd). On the other hand, for odd $N$, the two states
$\vert \psi (t_o)\rangle$, obtained measuring at $t=t_o$ the
internal state of the ion as $\vert - \rangle$, may be called
SU(2) Yurke-Stoler like coherent states, with a difference of
$\frac{\pi}{2}$ ($\frac{3 \pi}{2}$) in the relative quantum phase,
when $\frac{N+1}{2}$ is even (odd).

It is interesting to look at the states (\ref{s4.17}) and
(\ref{s4.18}) under another point of view \cite{jmomio}.

The angular momentum  $\hat{L}_z=2 \hbar \hat{J}_2$ can be express
in terms of the right and left circular quanta operators,
$\hat{n}_r=\hat{a}_r^{\dag} \hat{a}_r$ and
$\hat{n}_l=\hat{a}_l^{\dag} \hat{a}_l$ respectively, as follows
\cite{cohen}
\begin{equation}
L_z = \hbar \left( \hat{a}_r^{\dag} \hat{a}_r - \hat{a}_l^{\dag}
\hat{a}_l \right)
\end{equation}
It is not difficult to verify that the eigenstates $\vert n_l = N,
n_r =0 \rangle$ and $\vert n_l = 0, n_r =N \rangle$ of
$\hat{L}_z$, corresponding to its minimum ($- \hbar N$) and
maximum ($+ \hbar N$) eigenvalues respectively, coincide with the
SU(2) coherent states $\vert \tau = i , j=N/2 \rangle$ and $\vert
\tau = -i , j=N/2 \rangle$. Thus, the two contributing terms  in
the quantum superposition (\ref{s4.18}) are eigenstates of the
orbital angular momentum correspondent to its minimum and maximum
eigenvalue. In Fig.\ref{fig:9a} (a) we report the spatial
distribution  of the vibrational state (\ref{s4.18}) generated
with the help of our scheme. This plot clearly shows interference
effects which represent a signature of the quantum nature of this
superposition. Note that the spatial interference fringes are
sensitive to vibrational decoherence \cite{spatint} and can, thus,
be used to study the decoherence induced transition of the quantum
superpositions (\ref{s4.18}) into the correspondent statistical
mixture whose spatial distribution is shown in Fig.\ref{fig:9b}
(b). The vibrational state (\ref{s4.17}), generated in case of
even initial $N$, is on the contrary, a superposition of two
states describing ionic oscillations along the two orthogonal
directions $\bar{x}$ and $\bar{y}$ respectively.

\vspace{1cm}

{\it e) 1D Schr\"{o}dinger cats of bidimensionally confined ions}

We conclude this section considering an example of the procedure
for generating 1D cat states exploiting laser beams along the
$x-y$ plane. Usually in these schemes  the ion is supposed to be
confined in a highly anisotropic two-dimensional trap. To generate
a given state along the $x$ axis of the trap, it is assumed that
the ion is tightly bound in the $y$ direction and that the laser
along this axis is tuned to the atomic transition. The proposal we
wish to review \cite{gou1D} deals with the generation of even and
odd squeezed states of the form:
\begin{equation}
\vert \phi_{\pm} \rangle \propto \left(  \vert \alpha, \xi \rangle
\pm \vert -\alpha, \xi \rangle \right)
\end{equation} \label{s4.22}
where $\vert \alpha, \xi \rangle$ is the squeezed state defined by $\vert \alpha, \xi \rangle = D(\alpha) S(\xi)
\vert 0 \rangle$, with  $D(\alpha) = exp(\alpha \hat{a}^{\dag} - \alpha^* \hat{a})$ and $S(\xi)= exp(\xi
\hat{a}^{\dag 2} - \xi^* \hat{a}^2)$ displacement and \lq \lq squeeze\rq \rq operators respectively. Cat like
states as the one of equation (\ref{s4.22}) may be engineered as dark state solutions in presence of dissipation
due to spontaneous emission. Indeed it is not difficult to prove that even and odd squeezed states satisfy an
eigenvalue equation of the form
\begin{equation}
 \left( - g_1 e^{-i \phi_1} \hat{a}_x^2 - g_2 e^{-i \phi_2} \hat{a}^{\dag 2}_x- 2 g_0 e^{-i
\phi_0} \hat{a}_x^{\dag} \hat{a}_x \right) \hat{\rho} = \zeta
\hat{\rho}  \label{s4.24}
\end{equation}
In order to ensure that the steady state solution effectively
approaches a even or odd squeezed state of the form (\ref{s4.22}),
however, we need to assume that the initial state is  prepared
with a precise parity \cite{gou1D}. For example starting from the
Fock state $|0 \rangle$ or $| 1 \rangle$ one obtains even or odd
squeezed states respectively.

\section{Conclusive remarks}
Hamiltonian models describing the coupling between pseudo-spin and
bosonic dynamical variables are ubiquitous in physics. The reason
is that when one is faced with a complex problem like the
interaction of a material subsystem (atom, molecule, crystal,
etc.) with the radiation field (electromagnetic or elastic), such
models catch the essential ingredients of the physical problem and
often are exactly solvable. In this paper we have dealt with a
special class of such models, namely those describing the coupling
of a few-level atom with a pair of bosonic modes. We have shown
that these models are not oversimplified representations of given
physical situations only, but that, in fact, most of them have
been realized in the CQED and trapped ion experimental realms. Our
review discusses a wild collection of results achieved over the
last few years in both contexts elucidating that their occurrence
is chiefly due to the entanglement which gets established in the
tripartite system. In particular the possibility of inducing
entanglement dependent quantum correlation effects in the two
modes system is shown. This very attractive aspect as well as the
generation of many nonclassical bimodal states, clearly witness
the dynamical richness of the hamiltonian models which we have
focused on in this review paper.

\section*{Acknowledgements}
The authors acknowledge F.S. Persico for useful suggestions and
for reading part of the manuscript.

%%%%%%%%REFERENCES%%%%%%%%%%%%

%Fig.~\ref{fig:1}

\pagebreak

\begin{figure}
\caption{Energy level diagram of a three-level atom  (a) in the $\Lambda$ configuration; (b) in the $\Xi$ configuration and (c) in the $V$ configuration.}
  \vspace*{0.2cm}
  \label{fig1}
\end{figure}

\begin{figure}
\caption{Average photon number in mode 1 for $g_1=g_2=g$ and $n=20$}
  \vspace*{0.2cm}
  \label{fig2}
\end{figure}

\begin{figure}
\caption{Energy level diagram of a three-level atom  in the $\Lambda$ configuration. The detuning $\Delta$ is assumed large enough with respect to $E_3-E_1$ in order to adiabatically eliminate the level $E_2$.}
  \vspace*{0.2cm}
  \label{fig3}
\end{figure}

\begin{figure}
\caption{Energy level diagram of a three-level atom  interacting with the two modes of the field.}
  \vspace*{0.2cm}
  \label{fig4}
\end{figure}

\begin{figure}
\caption{Time evolution of the $\mu=1$ average photon number, for
$r_1=r_2=s=0$, corresponding to the initial condition (a) $n=20$
and (b) $n=21$. }
  \vspace*{0.2cm}
  \label{fig5}
\end{figure}

%%%%%%%%%%%%%%%%%%%%%%%%%%%%%%%%%%%%%%%%%%%%%%%%%%%%%%%%%%%%%%%%%%%%%%
%%%%%%%%%%% FIGURE IONI IN TRAPPOLA%%%%%%%%%%%%%%%%%%%%%%%%%%%%%%%%%

\begin{figure}
\caption{Schematic view of the axial section of a Paul trap}
  \vspace*{0.2cm}
  \label{fig:1}
\end{figure}

\begin{figure}
\caption{Raman excitation scheme coupling the electronic states
$\vert + \rangle and \vert - \rangle$ and the vibrational levels
through a third upper electronic level $\vert s \rangle$}
  \vspace*{0.2cm}
  \label{fig:2}
\end{figure}

\begin{figure}
\caption{ Vibrational Raman scheme leaving the electronic state
unchanged}
  \vspace*{0.2cm}
  \label{fig:3}
\end{figure}

\begin{figure}
\caption{Scheme for the measurement of the electronic state of the
ion by means of quantum jumps techniques}
  \vspace*{0.2cm}
  \label{fig:4}
\end{figure}

\begin{figure}
\caption{Time evolution of  $\langle \hat{J}_{1 }(t) \rangle$ for
initial  total number of vibrational quanta $N=20$ (gray line) and
$N=21$ (black line) }
  \vspace*{0.2cm}
  \label{fig:5}
\end{figure}

\begin{figure}
\caption{Probability distributions in the process of generating
the pair coherent state $\vert\alpha \beta ,q=2>_{pcs}$ from the
initial coherent state $ \vert\alpha ,\beta >$ with $|\alpha
|^2=2.5$ and $|\beta |^2=1.5$. The phase difference of the two
driving beams is $\phi =\pi ,$ and $\,\chi/(\Omega _{Lx}-\Omega
_{Ly})=0.0049$  }
  \vspace*{0.2cm}
  \label{fig:7}
\end{figure}

\begin{figure}
\caption{  Spatial distribution  of the superposition
$\frac{1}{\sqrt{2}} \left(  \vert n_l =N, n_r =0 \rangle + i \vert
n_l =0, n_r=N \rangle \right)$, for $N=21$, against $x'=
\frac{x}{\beta}$ and $y'= \frac{y}{\beta}$ with $\beta =
\sqrt{\frac{m \nu}{\hbar}}$  }
  \vspace*{0.2cm}
  \label{fig:9a}
\end{figure}

\begin{figure}
\caption{Spatial distribution  of the statistical mixture of the
two states $\vert n_l =N, n_r =0 \rangle $ and $\vert n_l =0, n_r
=N \rangle$,  for $N=21$, against $x'= \frac{x}{\beta}$ and $y'=
\frac{y}{\beta}$ with $\beta = \sqrt{\frac{m \nu}{\hbar}}$  }
  \vspace*{0.2cm}
  \label{fig:9b}
\end{figure}

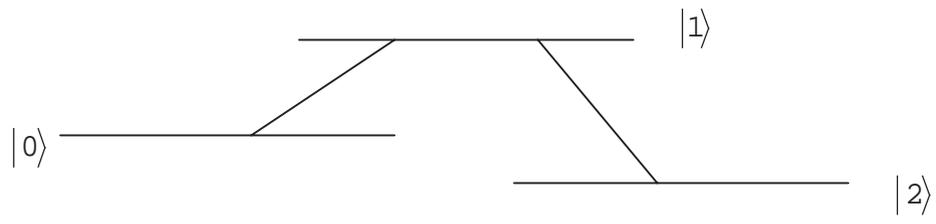
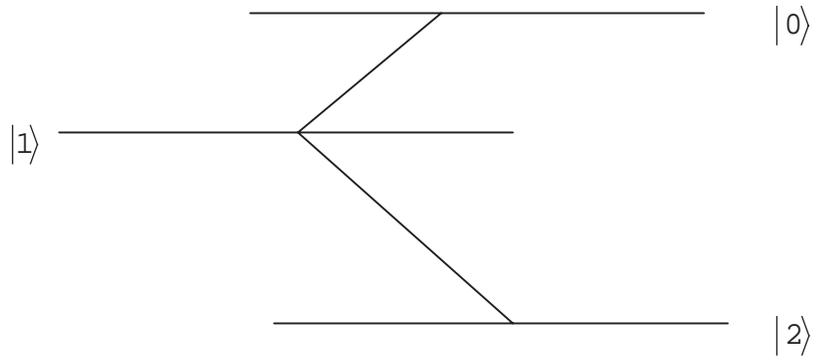
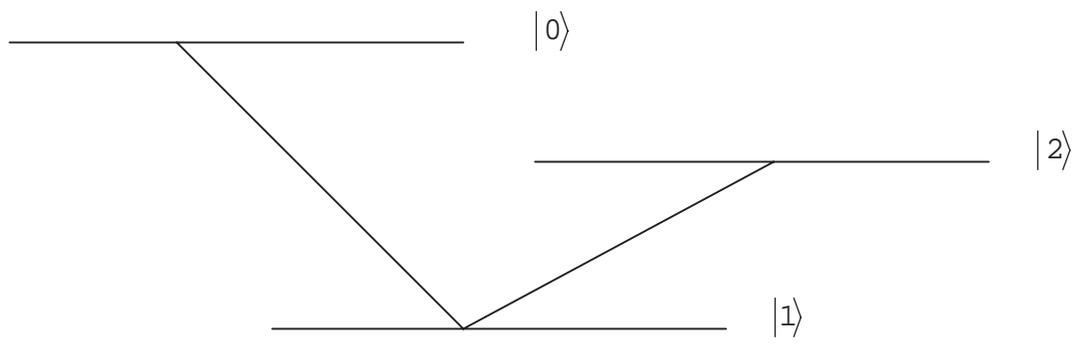
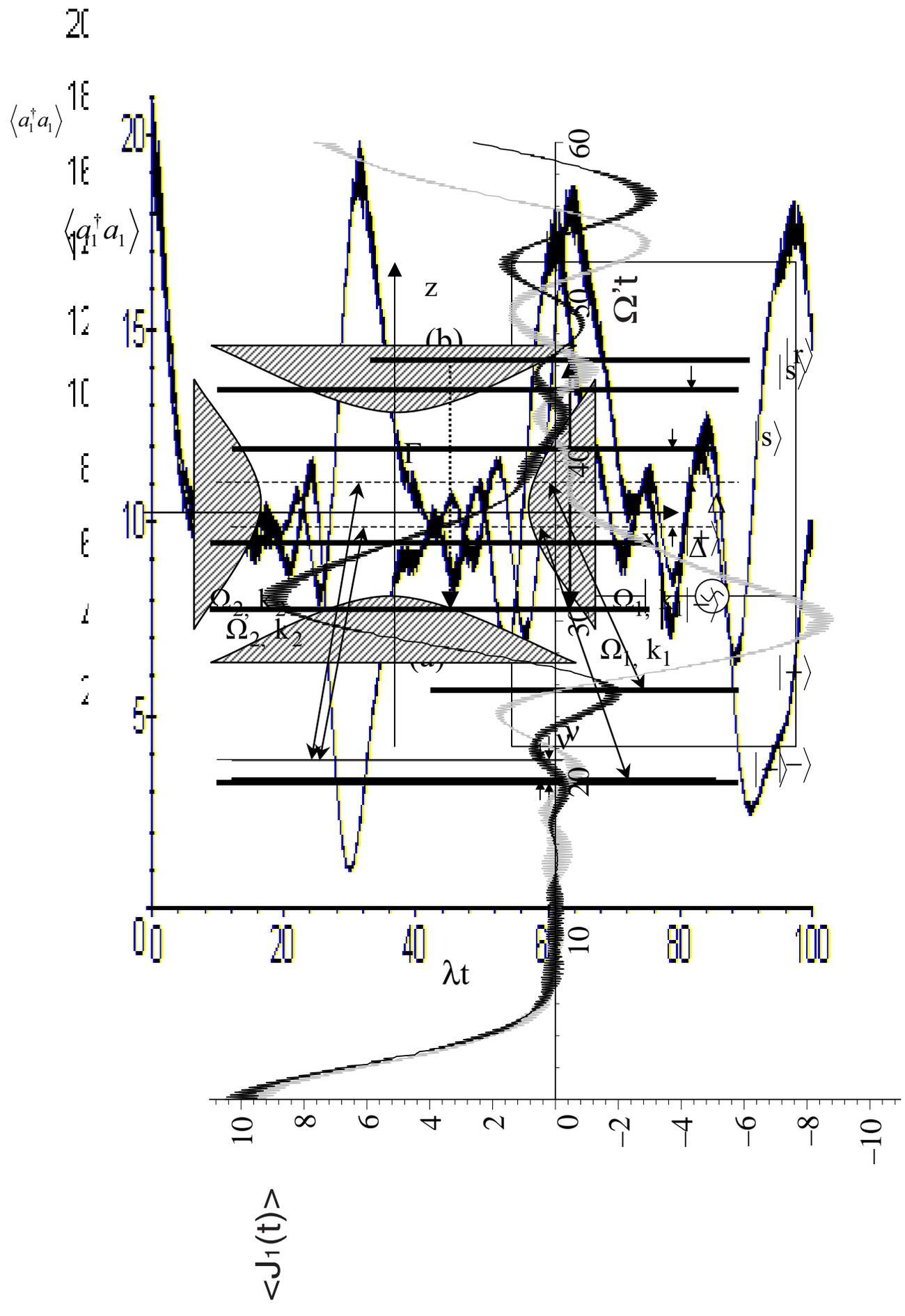
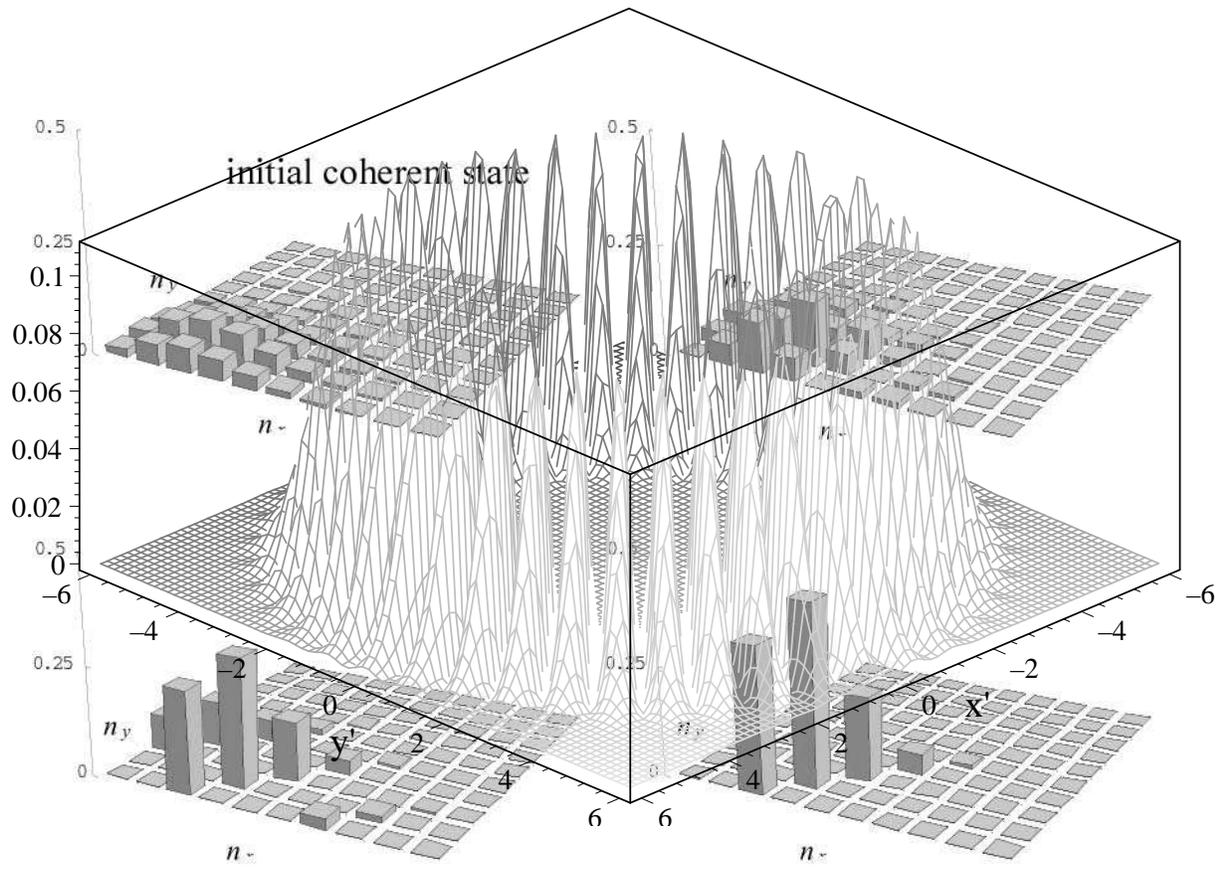
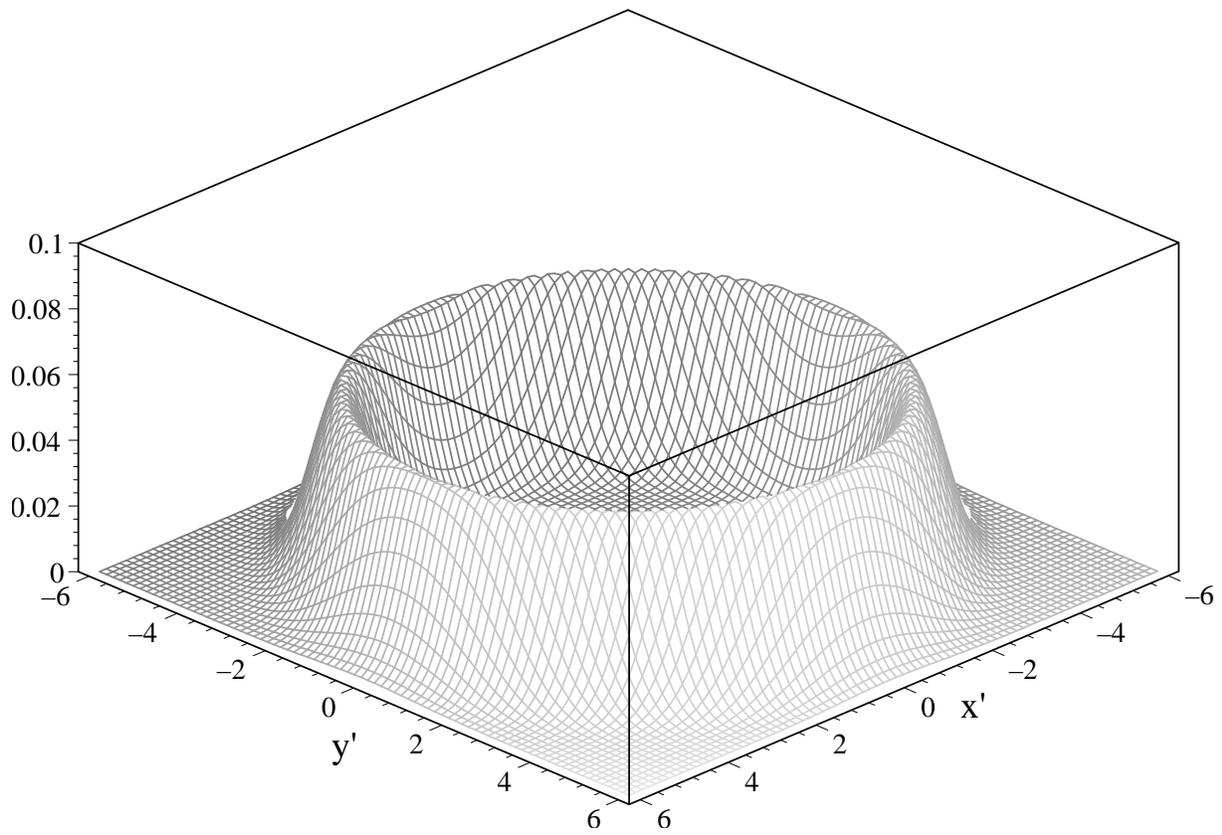
\end{document}